\pdfoutput=1
\documentclass[11pt,prd,preprintnumbers,nofootinbib,superscriptaddress,notitlepage]{revtex4-1}

\usepackage{graphics}
\usepackage{bm}
\usepackage{cancel}
\usepackage{color}
\usepackage{xcolor}
\usepackage{multirow}
\usepackage{slashed}
\usepackage{array}
\usepackage{amssymb}
\usepackage{graphicx}%Í¼Æ¬ºê°ü
\usepackage{mathrsfs}% »¨Ìå×ÖÄ¸
\usepackage{diagbox}% ±í¸ñÍ·Ð±Ïß
\usepackage{amsmath}%×Ó¹«Ê½
\usepackage{float}%´ò¿ªÓÐÂ·¾¶ÎÄ¼þºê°ü
\usepackage{extarrows}% ¼ýÍ·¹¦ÄÜ
\graphicspath{{figures/},{pics/}}
\usepackage[colorlinks,citecolor=blue,anchorcolor=red,menucolor=red,linkcolor=red,filecolor=red,runcolor=red,urlcolor=red,frenchlinks=red]{hyperref}
\usepackage{subfigure}

\makeatletter
\newcommand{\thickhline}{\noalign {\ifnum 0=`}\fi \hrule height 1pt\futurelet \reserved@a \@xhline}
\newcolumntype{"}{@{\hskip\tabcolsep\vrule width 1pt\hskip\tabcolsep}}                             
\makeatother

\def\pslash{\not{\hbox{\kern-4pt $p$}}}
\def\qslash{\not{\hbox{\kern-4pt $q$}}}
\def\lv{\not{\hbox{\kern-4pt $L$}}}
\def\lsim{\mathrel{\raise.3ex\hbox{$<$\kern-.75em\lower1ex\hbox{$\sim$}}}}
\def\gsim{\mathrel{\raise.3ex\hbox{$>$\kern-.75em\lower1ex\hbox{$\sim$}}}}
\def\ifmath#1{\relax\ifmmode #1\else $#1$\fi}

\def\slash{\!\!\!/}

\begin{document}
	\title{CP violation in lepton-number-conserving processes through heavy Majorana neutrinos at future lepton colliders}% Force line breaks with \\
	
	\author{Zhe Wang}
	\email{wzhe@mail.sdu.edu.cn}
	\affiliation{School of Physics, Shandong University, Jinan, Shandong 250100, China}
	
	\author{Xing-Hua Yang}
	\email{yangxinghua@sdut.edu.cn}
	\affiliation{School of Physics and Optoelectronic Engineering, Shandong University of Technology, Zibo, Shandong 255000, China}	
	
	\author{Xin-Yi Zhang}
	\email{xinyizhang@mail.sdu.edu.cn}
	\affiliation{School of Physics, Shandong University, Jinan, Shandong 250100, China}
	
	\date{\today}
	
	\begin{abstract}
         Small neutrino masses confirmed in the neutrino oscillation experiments indicate the need for new physics beyond the standard model. Seesaw mechanism is an interesting way to extend the standard model for explaining the neutrino masses. In a low-scale type-I seesaw mechanism, the tiny masses of neutrinos can be explained by heavy Majorana neutrino masses. Heavy Majorana neutrinos can lead to lepton-number-violating processes and the induced CP violation can contribute to the baryon asymmetry in the Universe. Heavy Majorana neutrinos can also lead to lepton-number-conserving processes and in this paper, we study the CP violation in lepton-number-conserving processes through heavy Majorana neutrinos at future lepton colliders. New possible observations of CP violation can also be connected to evidences of new physics beyond the standard model.
     \end{abstract}

	 % PACS, the Physics and Astronomy Classification Scheme.
	% 14.60.Pq Neutrino mass and mixing
	% 14.60.St Non-standard-model neutrinos, right-handed neutrinos, etc.
	
	%\keywords{Suggested keywords}%Use showkeys class option if keyword display desired
	
	\maketitle
\section{INTRODUCTION}\label{sec1}
In the standard model (SM), the neutrinos are strictly massless due to the absence of right-handed chiral states and the requirement of $SU(2)_L$ gauge invariance and renormalizability. The tiny masses of neutrinos confirmed in neutrino oscillation experiments show that the SM may not be complete. In fact, the SM has some open problems, such as the flavor puzzle, the existence of the dark matter (DM), the baryon asymmetry in the Universe. However, by introducing heavy Majorana neutrinos~\cite{Canetti:2012vf,Caputo:2018zky,Gninenko:2012anz}, we can solve some of them. A sector of Majorana neutrinos connected to the SM by mixing with the SM neutrinos could exhibit additional CP violation needed to explain the baryon asymmetry in the Universe. Heavy Majorana neutrinos can be used to explain the tiny masses of neutrinos via the interesting type-I seesaw mechanism~\cite{Mohapatra:1979ia,Minkowski:1977sc,Gell-Mann:1979vob}. 

In this work, we introduce three generations of right-handed heavy Majorana neutrinos in SM and the Dirac neutrino mass terms will be generated after spontaneous gauge symmetry breaking. The heavy Majorana neutrinos will lead to processes with violation of lepton number by two units $\Delta L=2$, such as the neutrinoless double-beta decay ($0\nu\beta\beta$)~\cite{Furry:1939qr,Elliott:2004hr}. It is an interesting process that the Majorana phase may induce additional CP violation effect. In previous works~\cite{Lu:2021vzj,Lu:2022pvw,Lu:2022wsm}, we have studied the lepton-number-violating (LNV) processes with heavy Majorana neutrinos and the induced CP violation. Not only that, the heavy Majorana neutrinos can also lead to lepton-number-conserving (LNC) processes, in this work, we study several interesting LNC processes caused by heavy Majorana neutrinos at future lepton colliders. The CP violation stems from the interference of contributions from different heavy Majorana neutrinos, and even one heavy Majorana neutrino can lead to CP violation in these processes by the intereference of contributions from the $s$-channel processes and the $t$-channel processes. We also investigate the prospects for searching for these heavy Majorana neutrinos at future lepton colliders like the Muon Collider (MuC)~\cite{MuonCollider:2022nsa} and the International Linear Collider (ILC)~\cite{Behnke:2013xla}. We analyse the processes $e^+e^-\rightarrow\bar{\nu_e}N_i\rightarrow \bar{\nu_e} e^-q\bar{q}^\prime$ at $e^+e^-$ collision with ILC running at 500 GeV, 1000 GeV, 3000 GeV and the processes $\mu^+\mu^-\rightarrow\bar{\nu_\mu}N_i\rightarrow \bar{\nu_\mu} \mu^-q\bar{q}^\prime$ at  $\mu^+\mu^-$ collision with MuC running at 3000 GeV and 10 TeV, where $N_i$ represent three generations of heavy Majorana neutrinos $N_1$, $N_2$, $N_3$.

This paper is organized as follows. In Section \ref{sec2}, we reviewed the model we used to describe heavy Majorana neutrinos. In Section \ref{sec3}, we analyse the CP violation in processes $e^+e^-\rightarrow\bar{\nu_e}N_i\rightarrow \bar{\nu_e} e^-q\bar{q}^\prime$ and $\mu^+\mu^-\rightarrow\bar{\nu_\mu}N_i\rightarrow \bar{\nu_\mu} \mu^-q\bar{q}^\prime$. The possibility for measuring CP violation at future lepton colliders is studied in Section \ref{sec4}. Finally, a short summary is given in Section \ref{sec5}. 
	\section{HEAVY MAJORANA NEUTRINOS BEYOND THE SM}\label{sec2}
The heavy neutrinos can only interact with the SM through mixing effects, which come from a mass matrix between the electroweak doublet neutrinos and Majorana neutrinos. In this work we extend the standard model with three right-handed heavy Majorana neutrinos. The Lagrangian of the model ~\cite{Mekala:2022cmm} we used in our process is given by:
\begin{equation}
	\label{1}
	\mathcal{L}=\mathcal{L}_{SM}+\mathcal{L}_{N}+\mathcal{L}_{WN\ell}+\mathcal{L}_{ZN\nu}+\mathcal{L}_{HN\nu} ,
\end{equation}
where $\mathcal{L}_{N}$ is a sum of kinetic and mass terms for heavy Majorana neutrinos:
\begin{equation}
	\label{2}
	\mathcal{L}_N=\frac{1}{2} \left(\bar{N_i}i\partial\!\!\!/N_i-m_{N_i}\bar{N_i}N_i\right),  
\end{equation}
where $i=1,2,3$, stand for three heavy Majorana neutrinos. The $\mathcal{L}_{WN\ell}$ corresponds to heavy neutrino interactions with a $W$ boson:  
\begin{eqnarray}
	\label{3}
	\mathcal{L}_{WN\ell}=-\frac{g}{\sqrt{2}}W^+_\mu\sum_{i=1}^{3}\sum_{\ell=e}^{\tau}\bar{N_i}R_{\ell i}^\ast\gamma^\mu P_L\ell^- + \text{h.c.},	
\end{eqnarray}
The $\mathcal{L}_{ZN\nu}$ to interactions with a $Z$ boson:
\begin{equation}
	\label{4}
	\mathcal{L}_{ZN\nu}=-\frac{g}{2\text{cos}\theta_{W}}Z_{\mu}\sum_{i=1}^{3}\sum_{\ell=e}^{\tau}\bar{N_i}R_{\ell i}^\ast\gamma^{\mu}P_L\nu_{l} + \text{h.c.},
\end{equation}
then the $\mathcal{L}_{HN\nu}$ to interactions with a Higgs boson:
\begin{eqnarray}
	\label{5}
	\mathcal{L}_{HN\nu}=-\frac{gm_N}{2M_W}h\sum_{i=1}^{3}\sum_{\ell=e}^{\tau}\bar{N_i}R_{\ell i}^\ast  P_L\nu_l + \text{h.c.},	
\end{eqnarray}
Finally we write the weak charged-current interaction Lagrangian as:
\begin{align}
	\label{6}
	\mathcal{L}_{\rm cc}
	= -\frac{g}{\sqrt{2}} W^{+}_{\mu} \sum_{m=1}^{3}\sum_{\ell=e}^{\tau}  V_{\ell m}^{\ast} \overline{\nu_{m}} \gamma^{\mu} P_{L} \ell^-
	- \frac{g}{\sqrt{2}} W^{+}_{\mu}  \sum_{i=1}^{3}\sum_{\ell=e}^{\tau} \bar{N_i}R_{\ell i}^{\ast}  \gamma^{\mu} P_{L} \ell^- + \text{h.c.}, \; 
\end{align}
Here $V_{\ell m}$ is the neutrino mixing matrix that can be measured from the neutrino oscillation experiments. The $R_{\ell i}$ indicates the mixing between heavy Majorana neutrinos and charged-leptons, which can be parameterized as~\cite{Xing:2007zj}.
\begin{equation}
	\label{7}
	R_{\ell i}=\left|R_{\ell i}\right|e^{i\phi_{\ell i}} , ~~~ \ell= e, \mu, \tau, ~~~ i=1, 2, 3 \; .
\end{equation}

The mixing between heavy Majorana neutrinos and $Z$ boson is different from those between heavy neutrinos and $W$ bosons, so the phases $\phi_{\ell i}$ should be set to different values. In this work we have at least 6 different phases as free parameters. For convenience, we set the phases between heavy neutrinos and $Z$ boson to $\phi_{\ell i}=0$, then we will set three free parameters of phases $\phi_{\ell 1}=\phi_a, \phi_{\ell 2}=\phi_b, \phi_{\ell 3}=\phi_c$ which are from mixing between heavy neutrinos and $W$ bosons to study each process and induced CP violation at ILC and MuC.

Now we give the mixing relations between the neutrino flavor eigenstates and mass eigenstates as follows\cite{Atre:2009rg} :
\begin{equation}
	\label{8}
	\nu_{\ell L} = \sum_{m=1}^{3} V_{\ell m} \nu_{m L} + \sum_{i=1}^{3} R_{\ell i} N^{c}_{iL} \; .
\end{equation}
In this work, we consider the case that the heavy Majorana neutrinos are nearly degenerate, so we set $\Gamma_{N_1}\approx\Gamma_{N_2}\approx\Gamma_{N_3}$, $m_{N_2}=m_{N_1}+\Gamma_{N_1}/2,m_{N_3}=m_{N_2}+\Gamma_{N_1}/2$ and we set $m_{N_1}$ in the range of 300 GeV-3000 GeV for ILC, 300 GeV-10 TeV for MuC according to Ref. \cite{Mekala:2022cmm} and Ref. \cite{Li:2023tbx}, for these ranges, at ILC, the parameters of the mixing between heavy neutrinos and leptons are set as $\left|R_{ei}\right|^2=\left|R_{\mu i}\right|^2=\left|R_{\tau i}\right|^2=0.0003$, and at MuC, the mixing parameters $\left|R_{\mu i}\right|^2$ are obviously dependent on $m_{N_1}$, we set the values of $\left|R_{\mu i}\right|^2$ according to the constraint in Fig.10 in Ref. \cite{Li:2023tbx}.

We simplify the widths of heavy Majorana neutrinos $\Gamma_{N_1},\Gamma_{N_2},\Gamma_{N_3}$ according to Refs.\cite{Si:2008jd,Atre:2009rg} in expression:

\begin{equation}
	\label{9}
\Gamma_{N_i}\simeq\left\{
\begin{aligned}
	&18\left(\frac{G_F^2m_{N_i}^5}{192\pi^3}\right) \Sigma_{\ell=e,\mu,\tau}\left|R_{\ell i}\right|^2,       m_{N_i}<M_W\\
	&{\cal A}\left(\frac{G_Fm_{N_i}^3}{8\sqrt2\pi}\right)\Sigma_{\ell=e,\mu,\tau}\left|R_{\ell i}\right|^2,{\cal A}=2(3)[4],m_{N_i}>M_W(M_Z)[M_H]  \\
\end{aligned}
\right.
\end{equation}
In the range of the mass $m_{N_1}$ we consider, we take ${\cal A}=4$, so that the widths can be simplified as:
\begin{equation}
	\label{10}
\Gamma_{N_i}\simeq\frac{G_Fm_{N_i}^3}{2\sqrt{2}\pi}\Sigma_{\ell=e,\mu,\tau}\left|R_{\ell i}\right|^2
\end{equation}	
We set the Cabibbo-Kobayashi-Maskawa (CKM) matrix as diagonal with unit entries for simplicity in our calculation. We put all these parameters in the model with $\text{F}_{\text{EYN}}\text{R}_{\text{ULES}}$~\cite{Alloul:2013bka}, the $\text{M}_{\text{ATHEMATICA}}$ package to calculate
Feynman rules associated with the Lagrangian of a given model, and use the model to generate the cross sections of the process with \textsf{MadGraph5\_aMC@NLO}~\cite{Alwall:2014hca}.

\section{$\text{CP}$ VIOLATION IN LEPTON-NUMBER-CONSERVING PROCESSES CONTRIBUTED BY HEAVY MAJORANA NEUTRINOS}\label{sec3}   

\begin{figure}[!htbp]
	\begin{center}
		\includegraphics[width=0.8\textwidth]{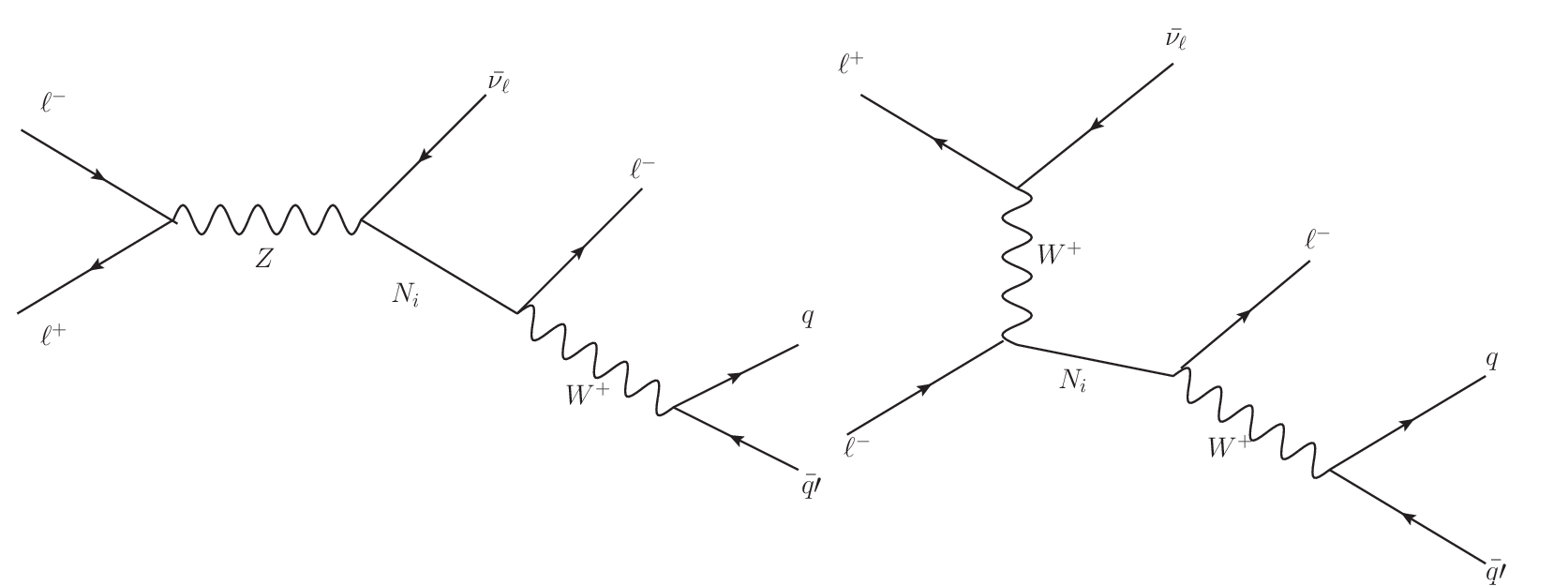}
		\caption{Feynman diagrams for process $\ell^+\ell^-\rightarrow \bar{\nu_\ell} N_i\rightarrow \bar{\nu_\ell} \ell^- q\bar{q}^\prime $}\label{fig1}
	\end{center}
\end{figure}

\begin{figure}[!htbp]
	\begin{center}
		\subfigure[]{\label{fig2a}
			\includegraphics[width=0.47\textwidth]{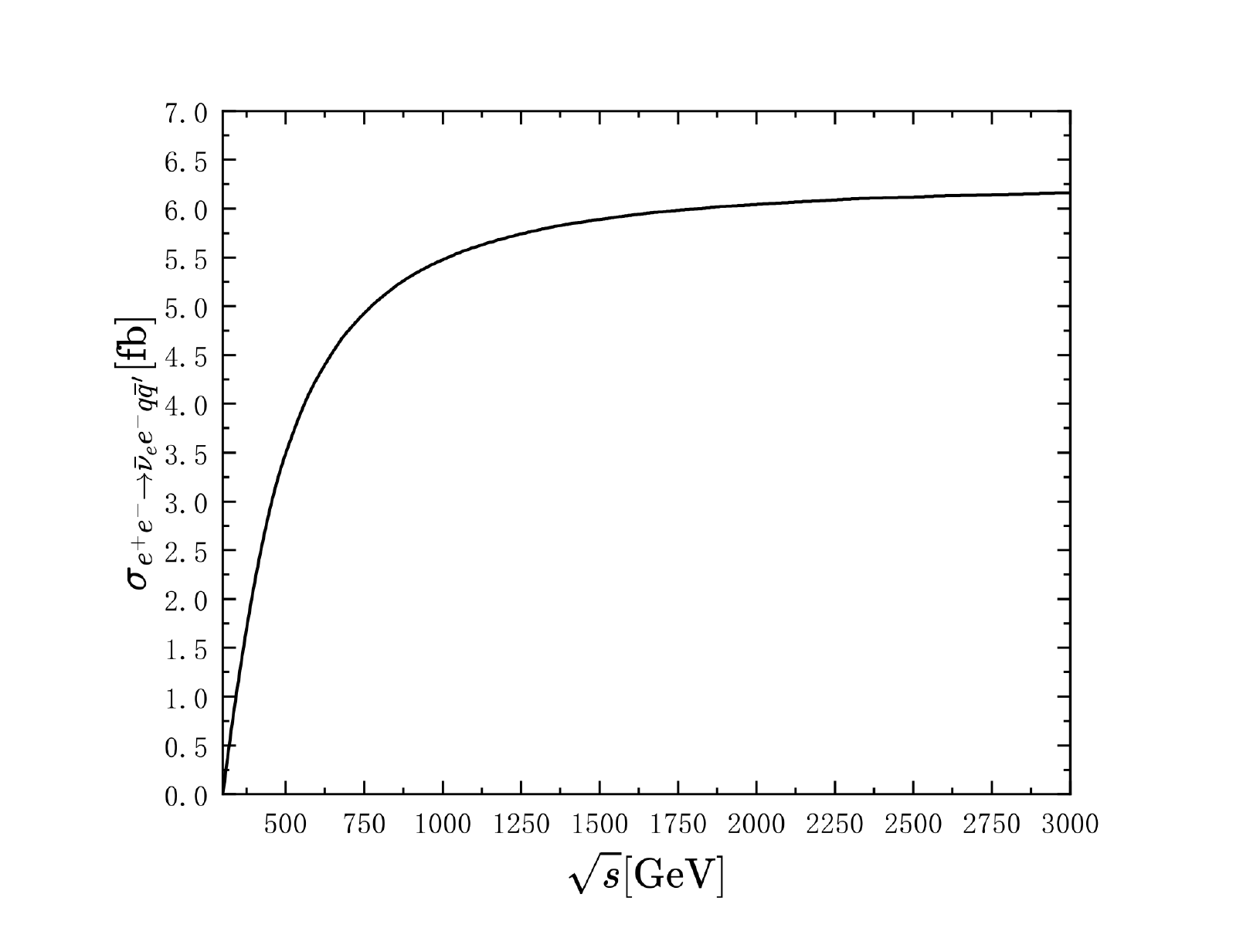} }
		\hspace{-0.5cm}~
		\subfigure[]{\label{fig2b}
			\includegraphics[width=0.47\textwidth]{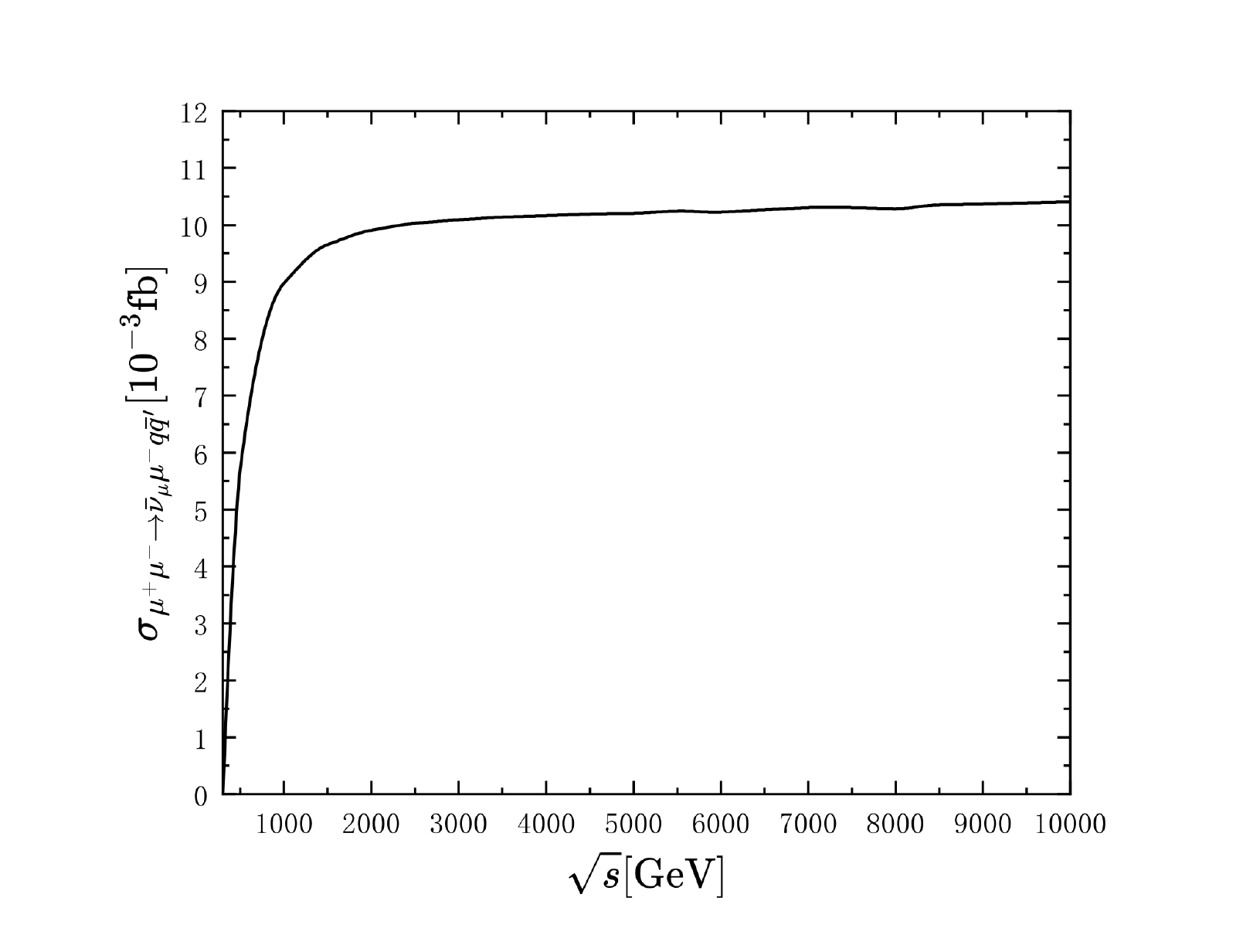} }
		\hspace{-0.5cm}~         	
		\caption{The total cross sections for (a) $e^+e^-\rightarrow\bar{\nu_e}e^-q\bar{q}^\prime$ and (b) $\mu^+\mu^-\rightarrow\bar{\nu_\mu}\mu^-q\bar{q}^\prime$ as versus the center-of-mass energy $\sqrt{s}$ with all CP phase set to $\phi_a=\phi_b=\phi_c=0$ and the Majorana neutrino mass set to $m_{N_1}=300$ GeV, and we set $\left|R_{\mu N}\right|^2=1\times10^{-5}$ in (b).  }\label{fig2}
	\end{center}
\end{figure}

The Feynman diagrams for our process are given in Fig.~\ref{fig1},
 where $\ell^\pm=e^\pm,\mu^\pm$. We can see there are two main diagrams, one is an $s$-channel diagram with $Z$ boson rare decay and the other one is a $t$-channel diagram with $W$ boson exchange. The $N_i$ respect three heavy Majorana neutrinos $N_1,N_2,N_3$. We take $q,q^\prime=c,s$ or $u,d$. The total cross section of this process can be expressed as
 
 \begin{equation}
	\label{11}
\sigma_{\ell^+\ell^-\rightarrow\bar{\nu_\ell}\ell^-q\bar{q\prime}}=\frac{1}{2s}\int\overline{\left|{\cal M}_{\ell^+\ell^-\rightarrow\bar{\nu_\ell}\ell^-q\bar{q\prime}}\right|^2}d{\cal L}_{ips4}
	\end{equation}
where the $\overline{\left|{\cal M}_{\ell^+\ell^-\rightarrow\bar{\nu_\ell}\ell^-q\bar{q\prime}}\right|^2}$ represents the squared matrix elements averaged (summed) over the initial (final) particles for the process with $d{\cal L}_{ips4}$ being the Lorentz invariant phase space of the four final particles. The $s$-channel process is dominant at the $Z$-pole (around the mass of the $Z$ boson), while for center-of-mass energy above the $Z$-pole, which belongs to the range we choose in this work, the $W$ exchange contribution is more important. The diagrams with heavy Majorana neutrinos $N_1$, $N_2$ and $N_3$ will give CP phase $\phi_{a}$, $\phi_{b}$ and $\phi_{c}$ in their $s$-channel diagrams which cause the CP violation of the total process $\ell^+\ell^-\rightarrow \bar{\nu_\ell}\ell^-jj$, and this CP phase does not exist in the $t$-channel diagrams. The total cross section is shown as the function of the center-of-mass energys $\sqrt{s}$ in Fig.~\ref{fig2}. The results in Fig.~\ref{fig2a} represent the case at ILC, and the results in Fig.~\ref{fig2b} 
corresponding to the case at MuC. They show that the total cross sections go up as the $\sqrt{s}$ increase, it means the contributions from $t$-channel are more important. 

\begin{figure}[!htbp]
	\begin{center}
		\subfigure[]{\label{fig3a}
			\includegraphics[width=0.47\textwidth]{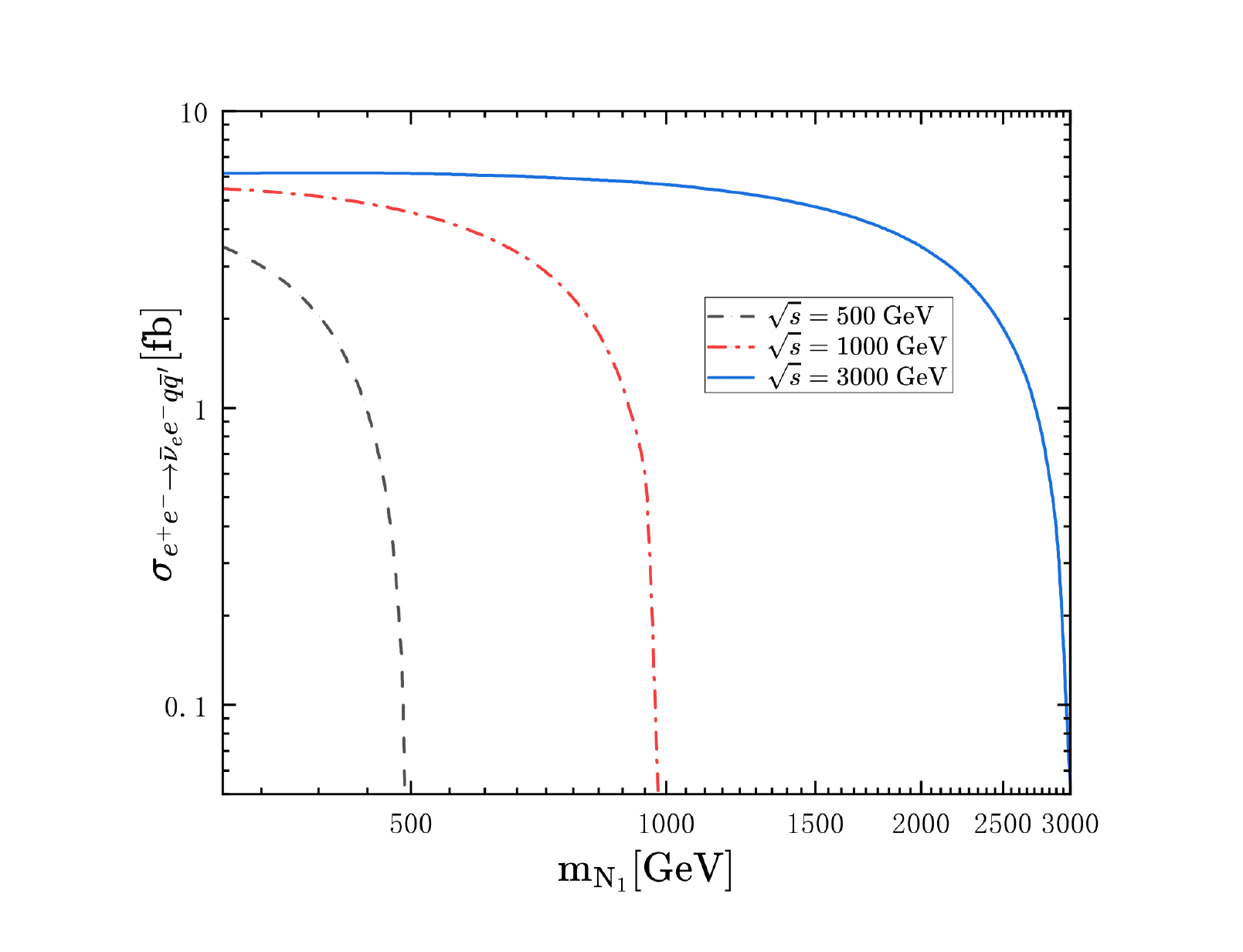} }
		\hspace{-0.5cm}~
		\subfigure[]{\label{fig3b}
			\includegraphics[width=0.47\textwidth]{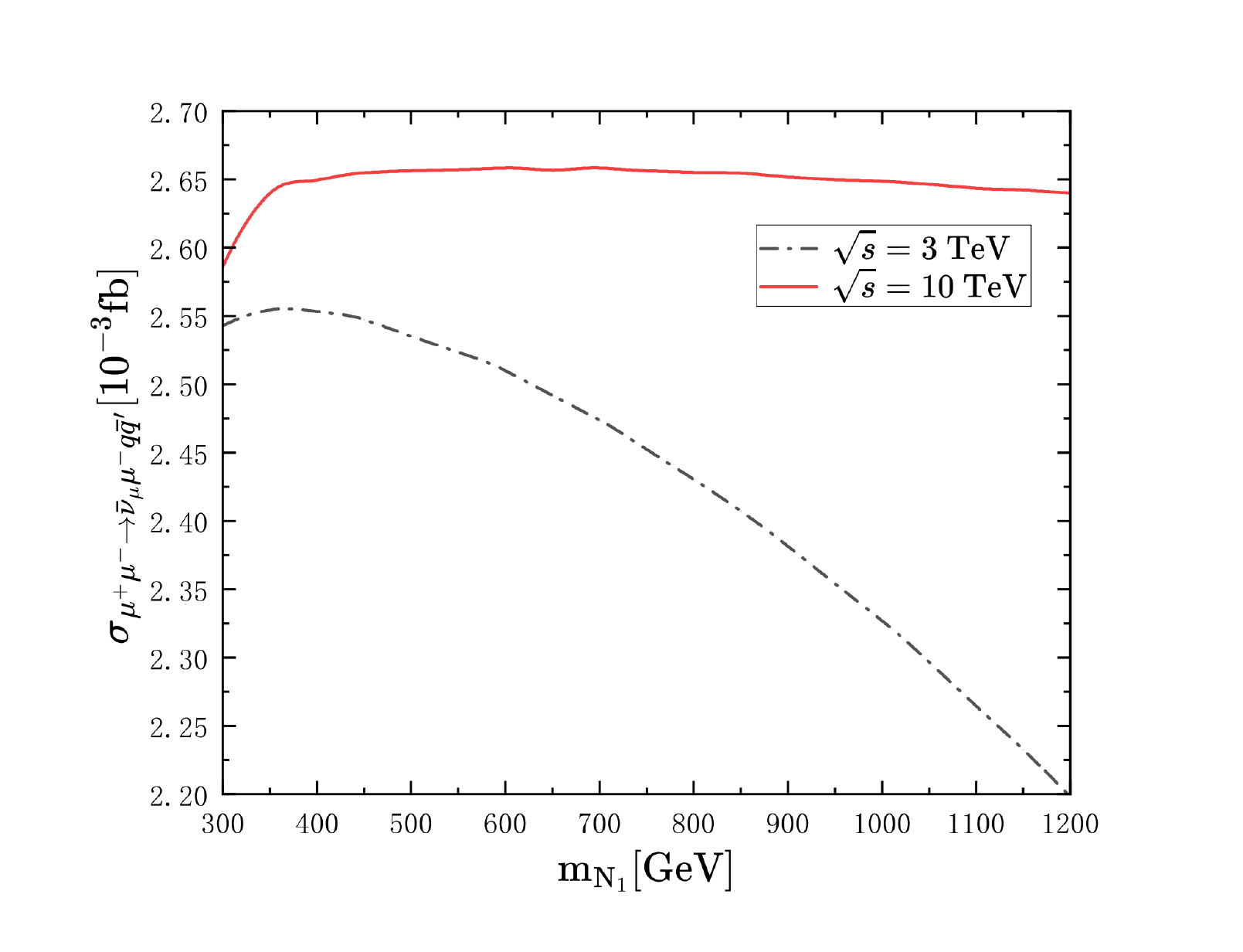} }
		\hspace{-0.5cm}~         	
		\caption{The total cross sections for (a) $e^+e^-\rightarrow\bar{\nu_e}e^-q\bar{q}^\prime$ at $\sqrt{s}=500$ GeV, $\sqrt{s}=1000$ GeV, $\sqrt{s}=3000$ GeV and (b) $\mu^+\mu^-\rightarrow\bar{\nu_\mu}\mu^-q\bar{q}^\prime$ at $\sqrt{s}=3000$ GeV, $\sqrt{s}=10$ TeV versus the Majorana neutrino mass $m_{N_1}$. We set $\left|R_{\mu N}\right|^2=5\times10^{-6}$ in (b). }\label{fig3}
	\end{center}
\end{figure}

The results for the total cross sections as the function of heavy Majorana neutrino mass $m_{N_1}$ at ILC and MuC are shown in Fig.~\ref{fig3a} and Fig.~\ref{fig3b} respectively.
 In Fig.~\ref{fig3a}, the dashed line represents the case $\sqrt{s}=500$ GeV at ILC, and the dash-dotted line represents $\sqrt{s}=1000$ GeV, full line represents $\sqrt{s}=3$ TeV. In Fig.~\ref{fig3b}, the dash-dotted line represents the case $\sqrt{s}=3$ TeV at MuC, and the full line represents $\sqrt{s}=10$ TeV. We can see that the cross section decrease quickly as the $m_{N_1}$ increase. The difference between the rates of $\ell^+\ell^-\rightarrow\bar{\nu_\ell}\ell^-q\bar{q\prime}$ and $\ell^-\ell^+\rightarrow\nu_\ell\ell^+\bar{q}q\prime$, where $\ell =e,\mu$ may induce the $\text{CP}$ asymmetry, which can be defined as 
\begin{equation}
	\label{12}
{\cal A}_{CP}=\frac{\sigma_{\ell^+\ell^-\rightarrow\bar{\nu_\ell}\ell^-q\bar{q\prime}}-\sigma_{\ell^-\ell^+\rightarrow\nu_\ell\ell^+\bar{q}q\prime}}{\sigma_{\ell^+\ell^-\rightarrow\bar{\nu_\ell}\ell^-q\bar{q\prime}}+\sigma_{\ell^-\ell^+\rightarrow\nu_\ell\ell^+\bar{q}q\prime}},
\end{equation}
As mentioned before, we have three CP phases $\phi_a,\phi_b,\phi_c$ as free parameters, they will cause CP violation in the processes $\ell^+\ell^-\rightarrow\bar{\nu_\ell}\ell^-q\bar{q\prime}$, for the case that if there is only one generation of heavy neutrinos, its $s$-channel diagram will give a CP phase but the $t$-channel diagram will not, so that this heavy Majorana neutrino will cause CP violation in this proces. We study the CP violation for cases that there is only one heavy Majorana neutrino, two generations of heavy Majorana neutrinos, and three generations of heavy Majorana neutrinos respectively. The results are in following pictures.
\begin{figure}[!htbp]
	\begin{center}
		\subfigure[]{\label{fig4a}
			\includegraphics[width=0.47\textwidth]{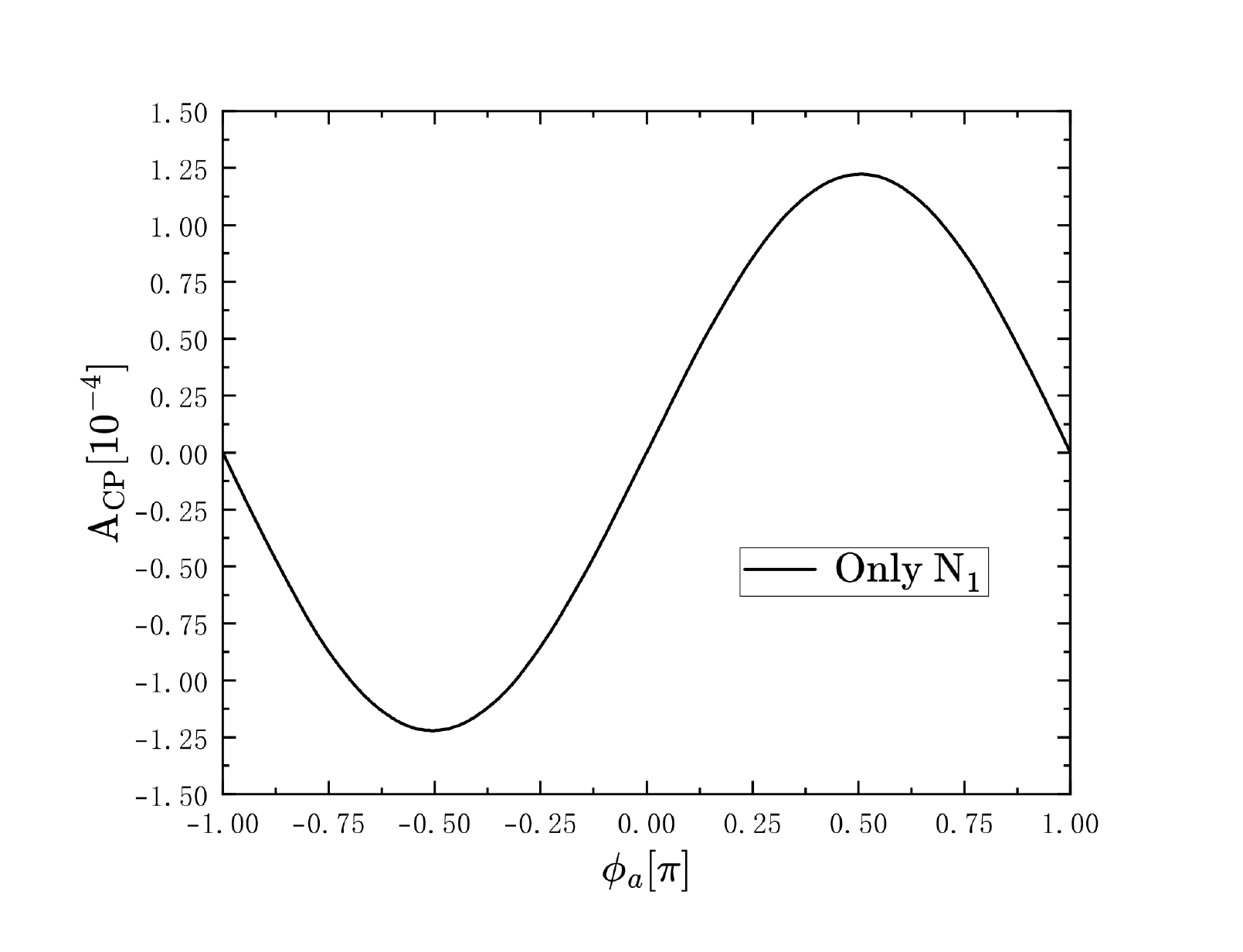} }
		\hspace{-0.5cm}~
		\subfigure[]{\label{fig4b}
			\includegraphics[width=0.47\textwidth]{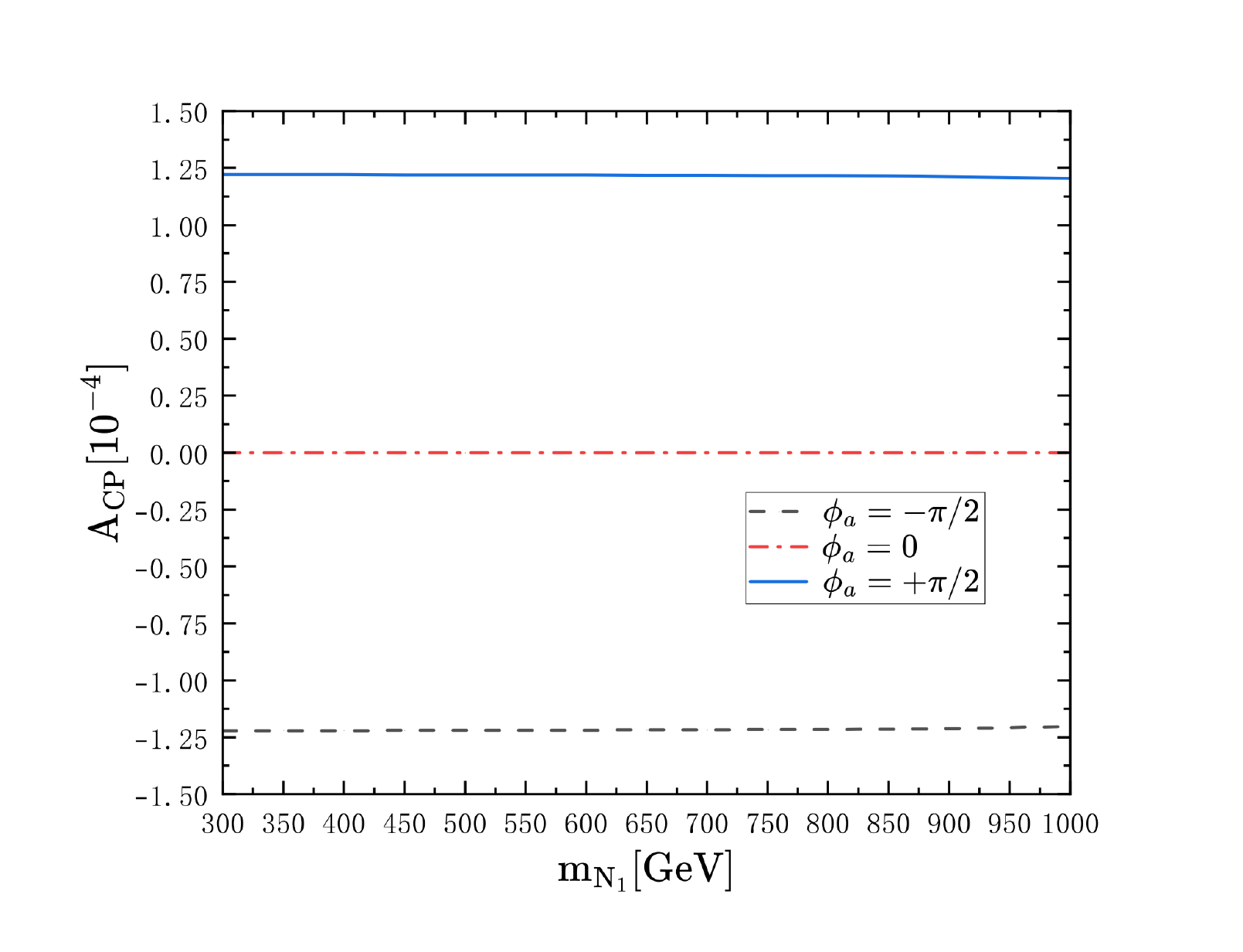} }
		\hspace{-0.5cm}~
        \subfigure[]{\label{fig4c}
     	\includegraphics[width=0.47\textwidth]{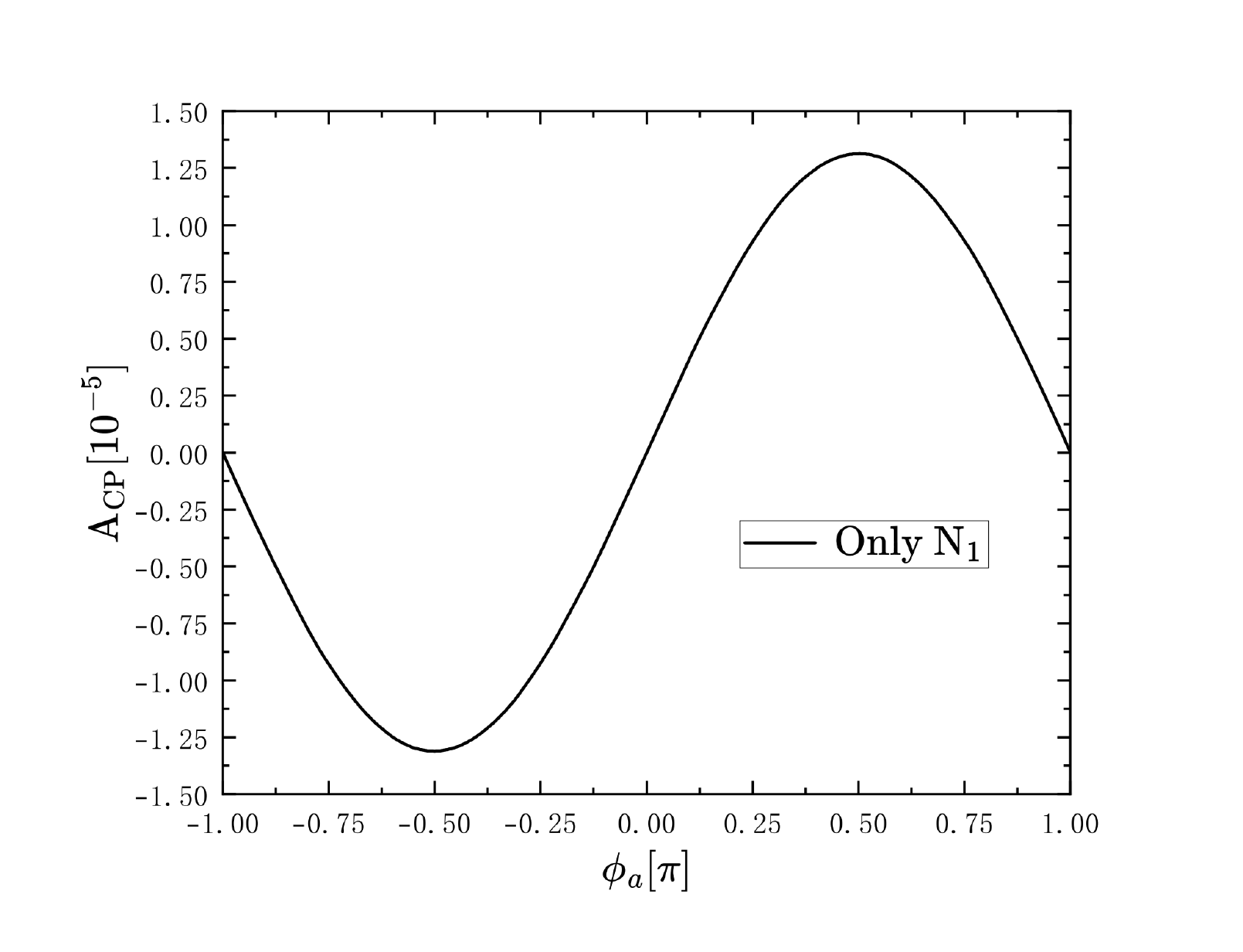} }	
		\hspace{-0.5cm}~
        \subfigure[]{\label{fig4d}
    	\includegraphics[width=0.47\textwidth]{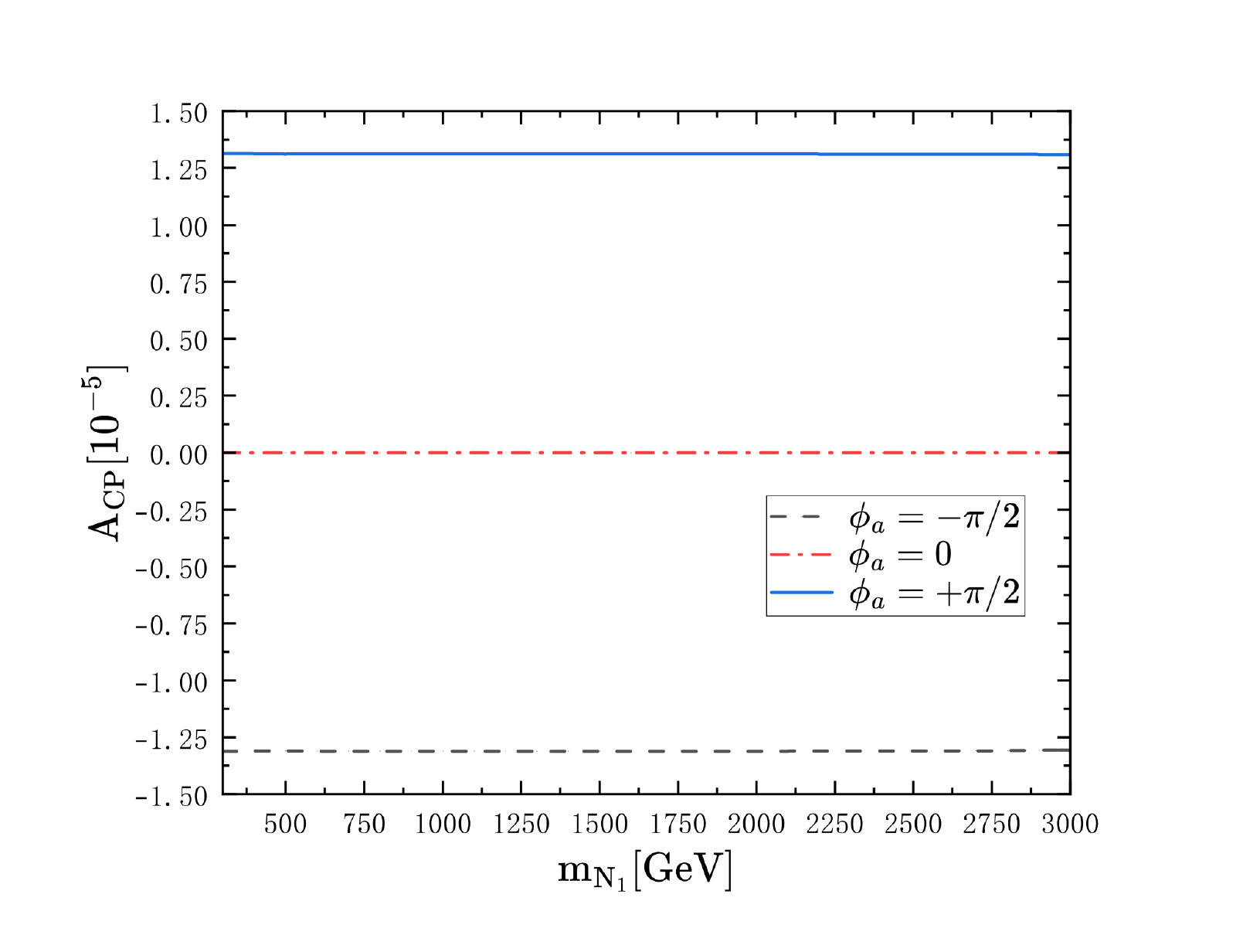} }          	
		\caption{CP violation for (a),(b) $e^+e^-\rightarrow\bar{\nu_e}e^-q\bar{q}^\prime$ and (c),(d) $\mu^+\mu^-\rightarrow\bar{\nu_\mu}\mu^-q\bar{q}^\prime$ with only one generation of heavy Majorana neutrino $N_1$ versus CP phase $\phi_a$ and Majorana neutrino mass $m_{N_1}$ where in (a) and (c) we set $m_{N_1}=300$ GeV. In (a),(b) we set $\sqrt{s}=1000$ GeV, $\left|R_{\mu i}\right|^2=3\times10^{-4}$. In (c), (d) we set $\sqrt{s}=3000$ GeV, and we set $\left|R_{\mu i}\right|^2=1\times10^{-5}$ in (c), $\left|R_{\mu i}\right|^2=4\times10^{-6}$ in (d).}\label{fig4}
	\end{center}
\end{figure}  
We show the results of CP violation as functions of mass $m_{N_1}$ and CP phase $\phi_a$ for case with only one generation of heavy Majorana neutrino here. Fig.~\ref{fig4a} and Fig.~\ref{fig4b} represent ILC case, Fig.~\ref{fig4c} and Fig.~\ref{fig4d} represent MuC case. We set $m_{N_1}=300$ GeV, $\sqrt{s}=1000$ GeV when we study the CP violation as the function of CP phases for $e^+e^-$ case in Fig.~\ref{fig4a}, and for $\mu^+\mu^-$ case, we set $\left|R_{\mu i}\right|^2=1\times10^{-5}$ in Fig.~\ref{fig4c}. When we study the CP violation as the function of $m_{N_1}$, as the mixing parameter $\left|R_{\mu i}\right|^2$ is obviously dependent on $m_{N_1}$ in $\mu^+\mu^-$ case, we set $\left|R_{\mu i}\right|^2=4\times10^{-6}$ for $\mu^+\mu^-$ case in Fig.~\ref{fig4d} and $\left|R_{\mu i}\right|^2=3\times10^{-4}$ in Fig.~\ref{fig4b} at the range 300 GeV $<m_{N_1}<$ 3000 GeV for convenience in this paper. From Fig.~\ref{fig4b} and Fig.~\ref{fig4d}, we can find that the CP violation is independent of the heavy Mjaorana neutrino mass $m_{N_1}$, in Fig.~\ref{fig4a} and Fig.~\ref{fig4c} we can see that the CP violation is related to CP phase, and the maximum value can reach near $1.25\times 10^{-4}$ at $\phi_a=\pm\pi/2$. This shows that only one generation of Majorana neutrino can lead to CP violation in LNC process even though it is small, which is different from that in LNV process, and the results for $\mu^+\mu^-$ case is not the same as that in $e^+e^-$ case, the function have the same shape but the values in $\mu^+\mu^-$ case are much smaller than those in $e^+e^-$, the  differences come from the different value of mixing $\left|R_{\mu i}\right|^2$ we take at ILC and MuC. 

\begin{figure}[!htbp]
	\begin{center}
		\subfigure[]{\label{fig5a}
			\includegraphics[width=0.47\textwidth]{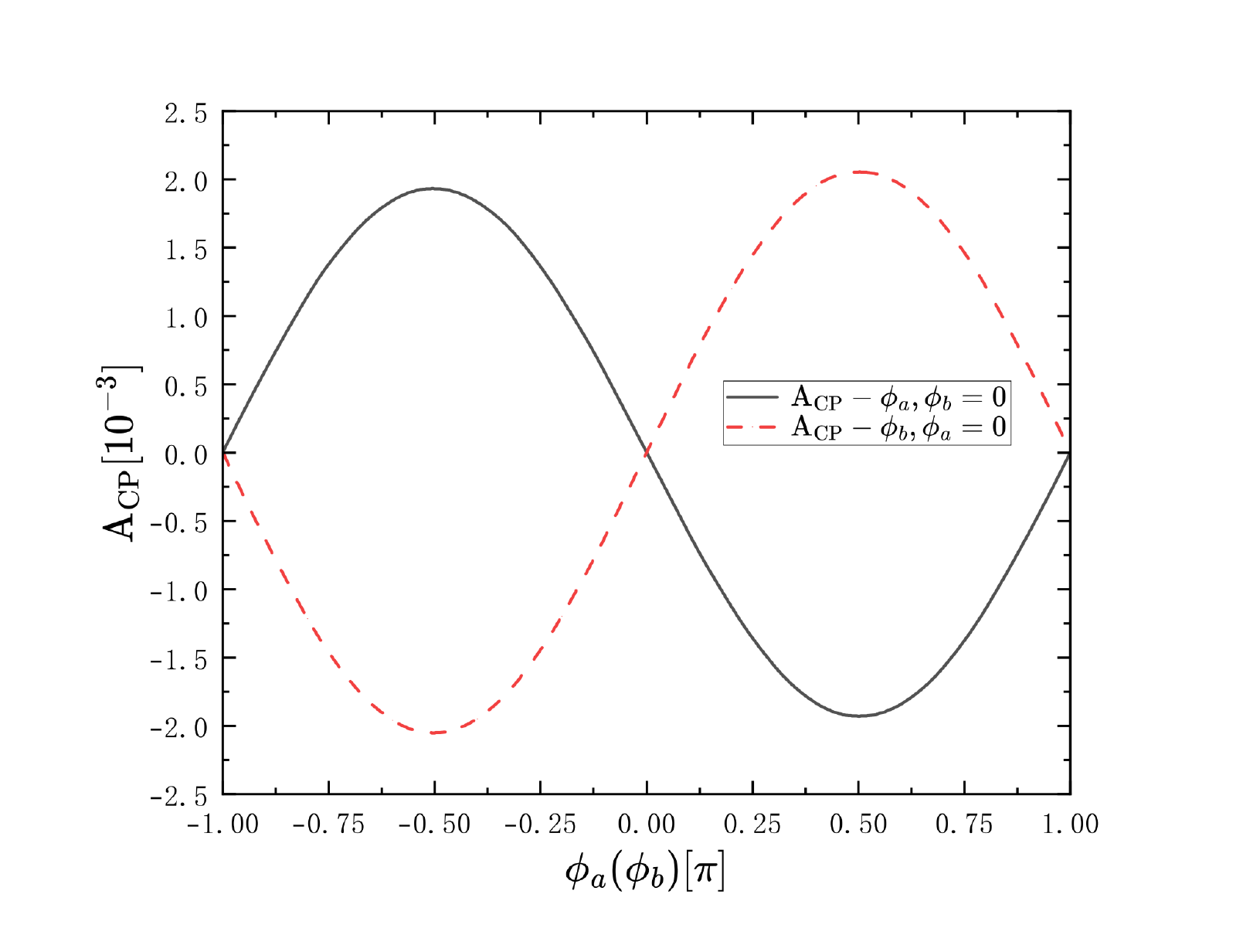} }
		\hspace{-0.5cm}~
		\subfigure[]{\label{fig5b}
			\includegraphics[width=0.47\textwidth]{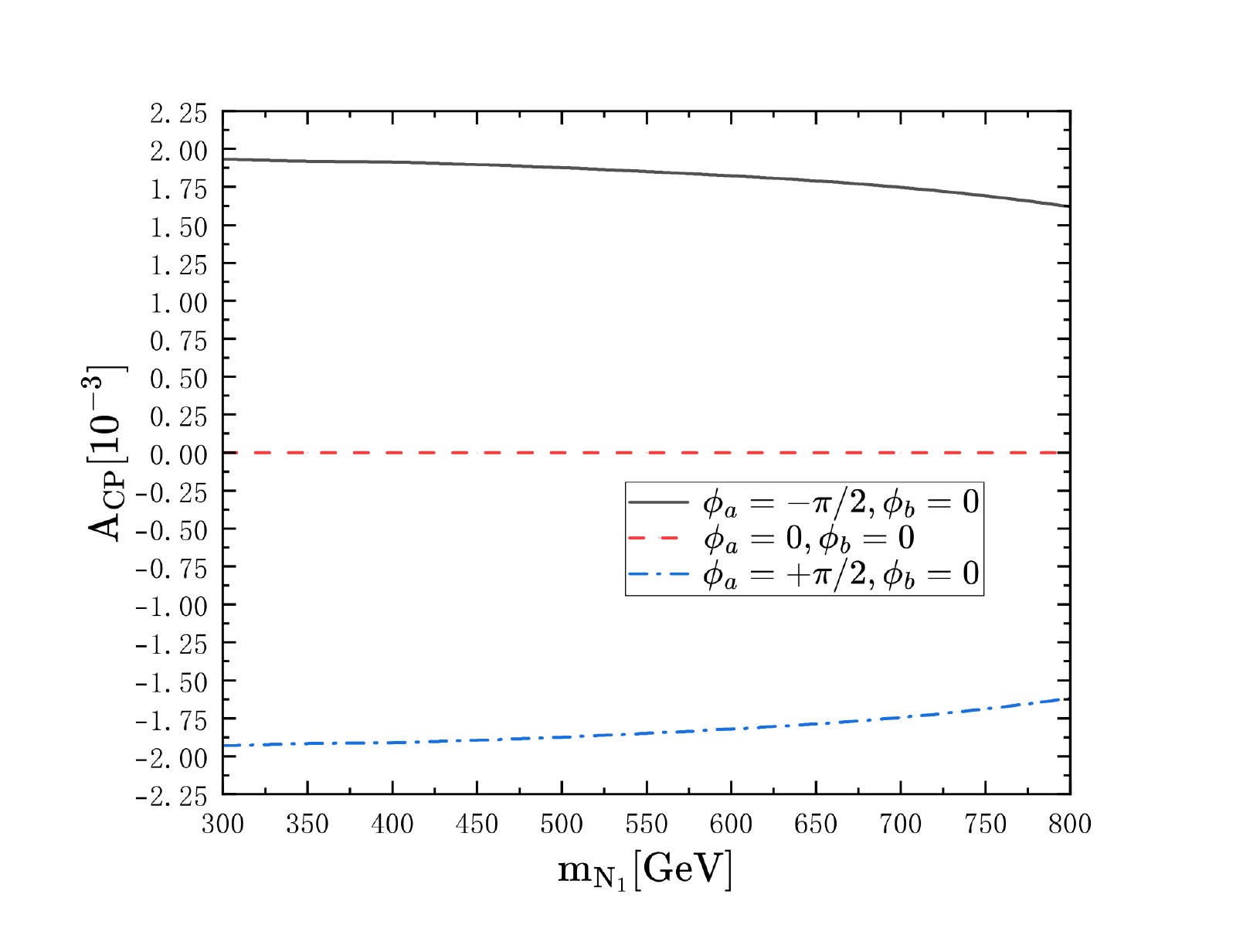} }
		\hspace{-0.5cm}~
		\subfigure[]{\label{fig5c}
			\includegraphics[width=0.47\textwidth]{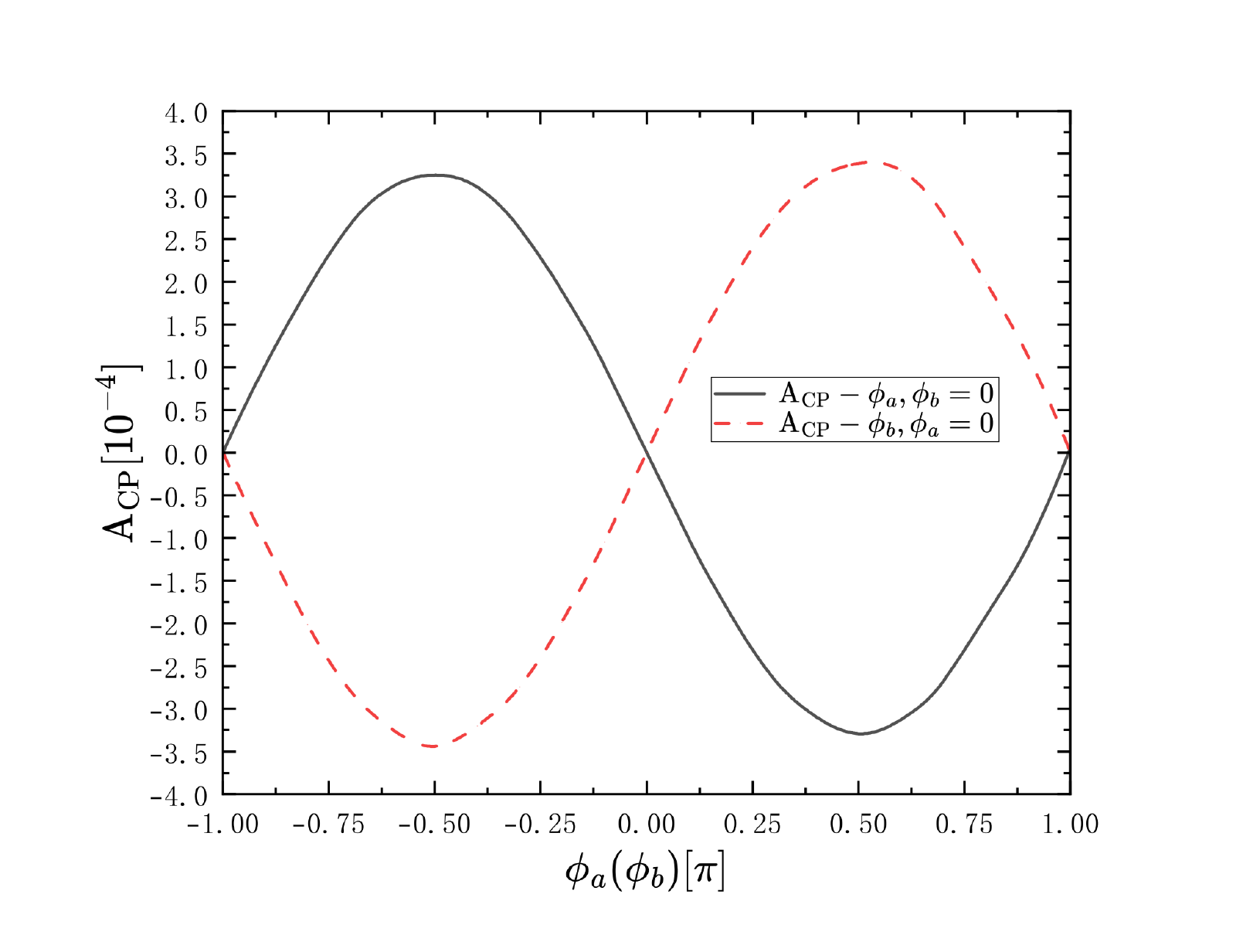} }	
		\hspace{-0.5cm}~
		\subfigure[]{\label{fig5d}
			\includegraphics[width=0.47\textwidth]{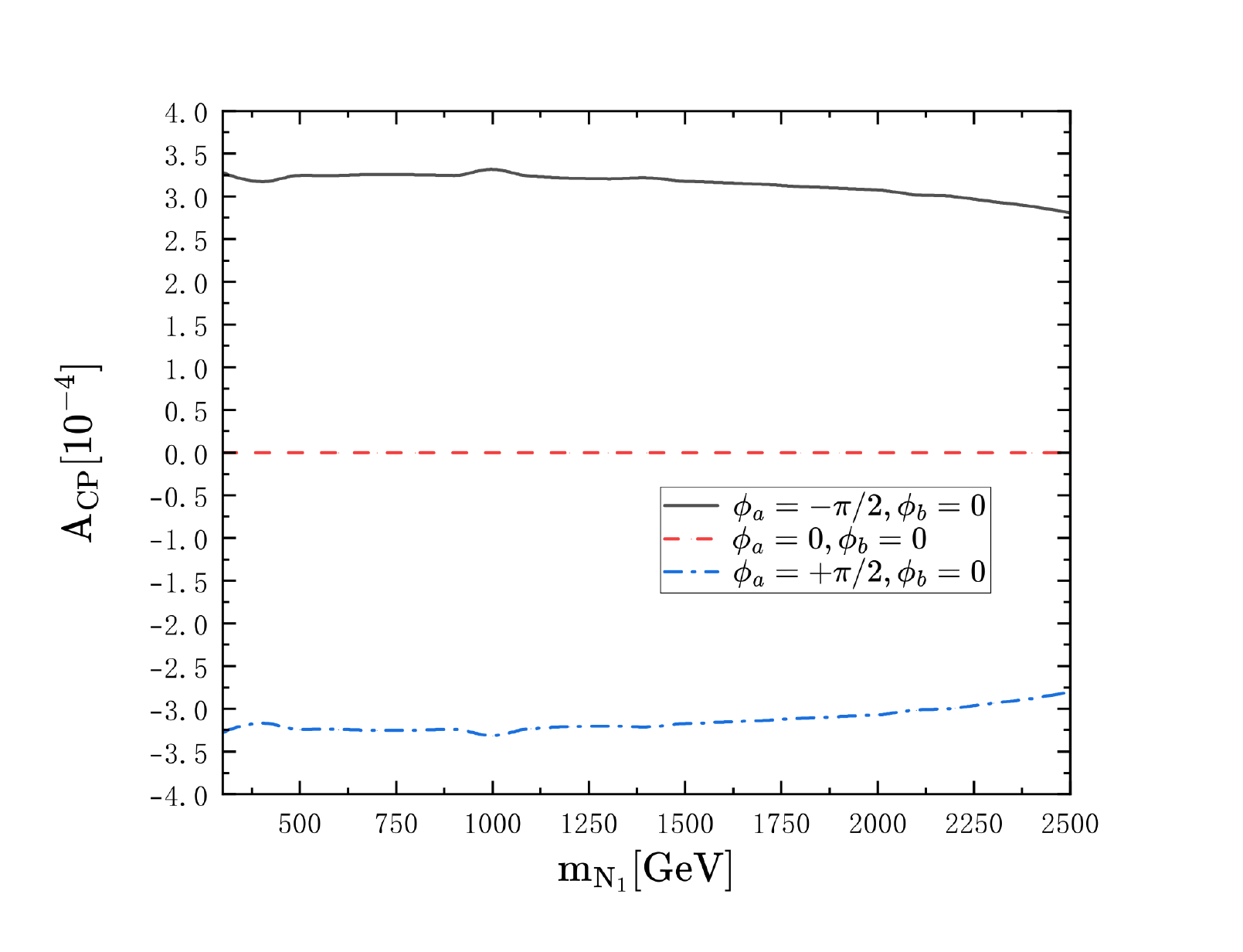} }          	
		\caption{CP violation for (a),(b) $e^+e^-\rightarrow\bar{\nu_e}e^-q\bar{q}^\prime$  and (c),(d) $\mu^+\mu^-\rightarrow\bar{\nu_\mu}\mu^-q\bar{q}^\prime$ with two generations of heavy Majorana neutrinos $N_1,N_2$ versus CP phase $\phi_a,\phi_b$ and Majorana neutrino mass $m_{N_1}$. In (a) and (c) we set $m_{N_1}=300$ GeV. In (a),(b) $\sqrt{s}=1000$ GeV, $\left|R_{\mu_{N_i}}\right|^2=3\times10^{-4}$. In (c), (d) we set $\sqrt{s}=3000$ GeV, and we set $\left|R_{\mu_{N_i}}\right|^2=1\times10^{-5}$ in (c), $\left|R_{\mu_{N_i}}\right|^2=4\times10^{-6}$ in (d).}\label{fig5}
	\end{center}
\end{figure}

The results with two generations of Majorana neutrinos $N_1$ and $N_2$ are shown in Fig.~\ref{fig5}.
 Fig.~\ref{fig5a} and Fig.~\ref{fig5c} show that the influences on CP violation from different CP phase $\phi_a$ and $\phi_b$, and when there are two generations of heavy Majorana neutrinos, the maxium value of CP violation will reach the order of magnitude of $1\times 10^{-3}$. However, from Fig.~\ref{fig5b} and Fig.~\ref{fig5d} we can see when there are two heavy Majorana neutrinos, the CP violation will be influenced by $m_{N_1}$.

\begin{figure}[!htbp]
	\begin{center}
		\subfigure[]{\label{fig6a}
			\includegraphics[width=0.47\textwidth]{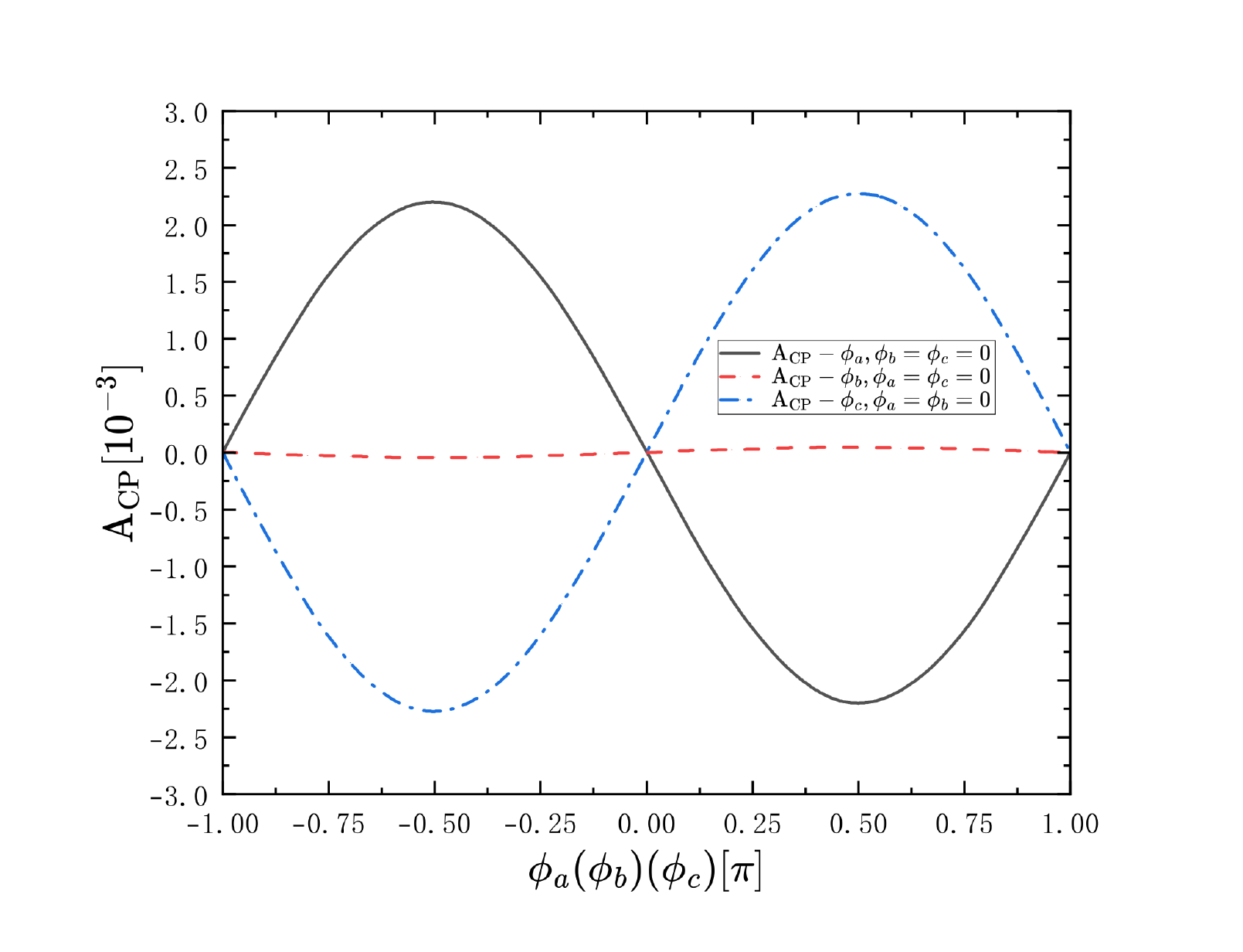} }
		\hspace{-0.5cm}~
		\subfigure[]{\label{fig6b}
			\includegraphics[width=0.47\textwidth]{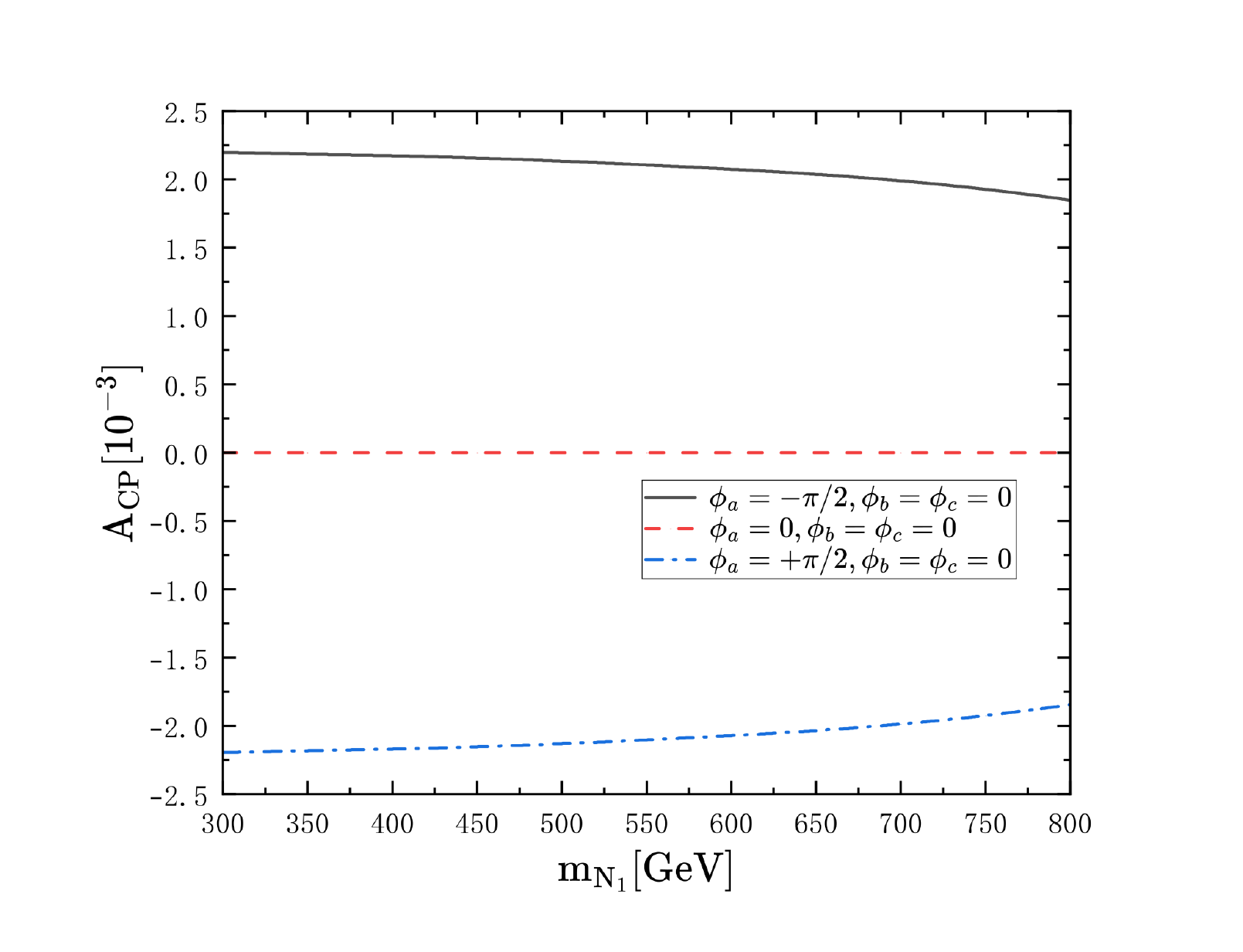} }
		\hspace{-0.5cm}~
		\subfigure[]{\label{fig6c}
			\includegraphics[width=0.47\textwidth]{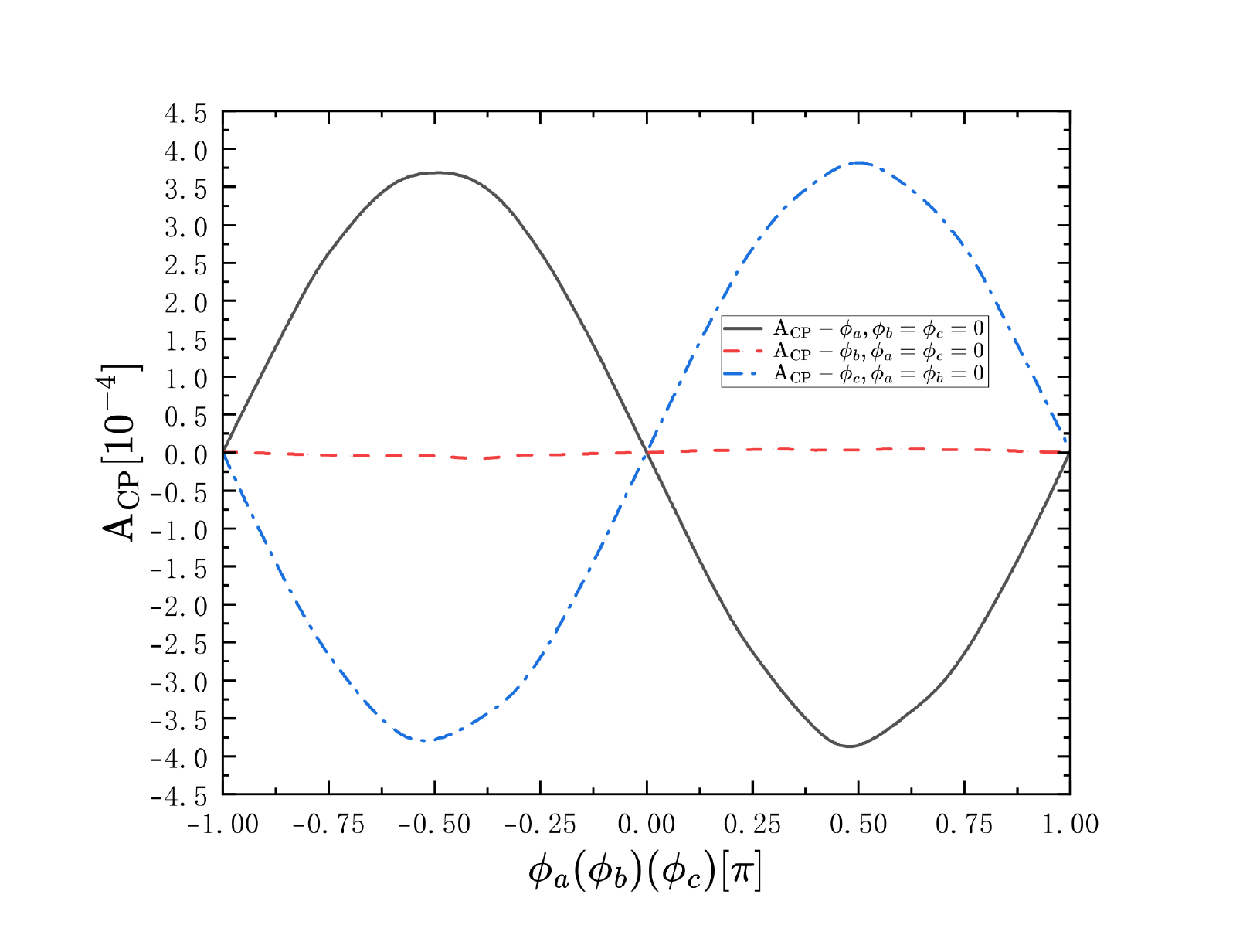} }	
		\hspace{-0.5cm}~
		\subfigure[]{\label{fig6d}
			\includegraphics[width=0.47\textwidth]{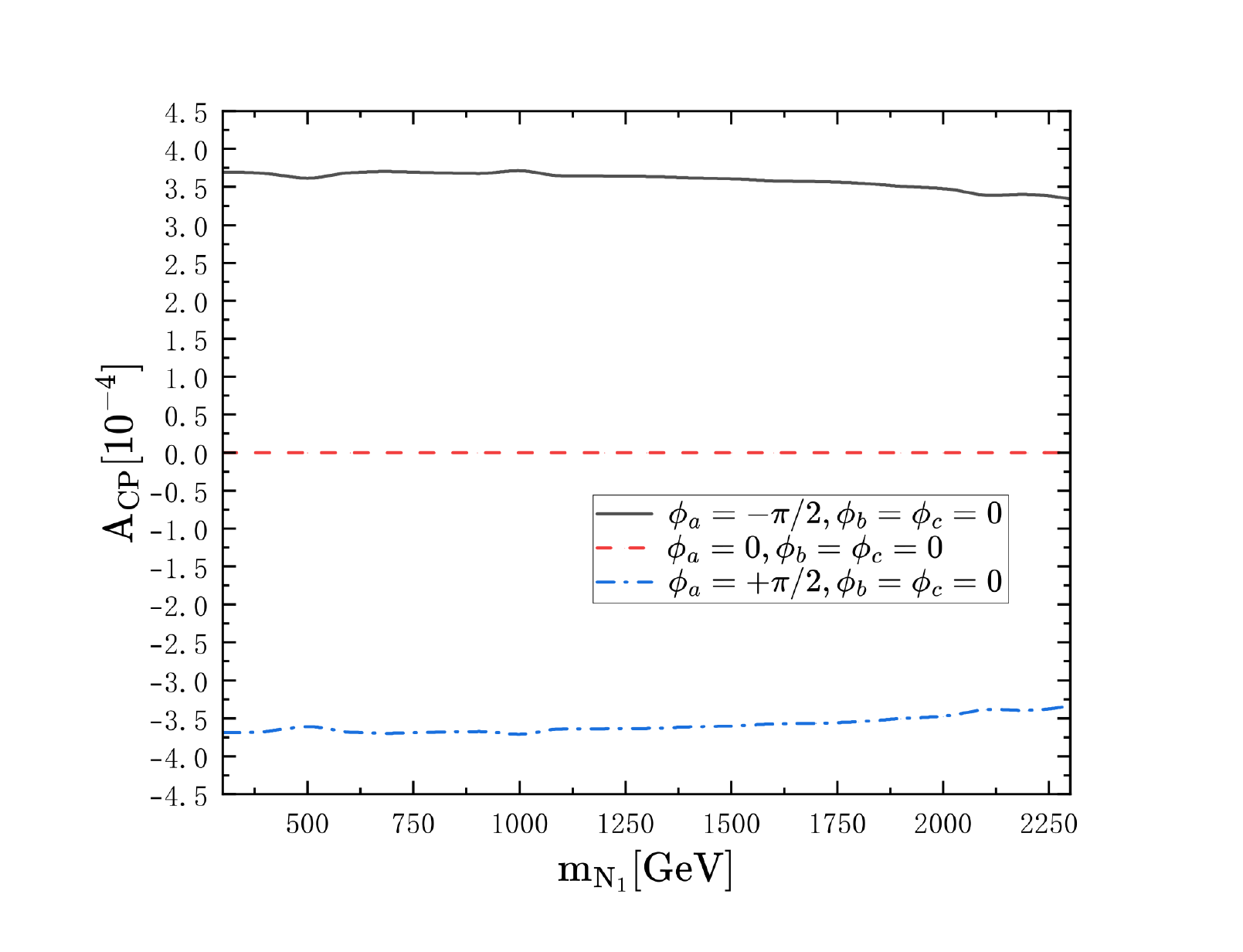} }          	
		\caption{CP violation for (a),(b) $e^+e^-\rightarrow\bar{\nu_e}e^-q\bar{q}^\prime$ and (c),(d) $\mu^+\mu^-\rightarrow\bar{\nu_\mu}\mu^-q\bar{q}^\prime$ with three generations of heavy Majorana neutrinos $N_1,N_2,N_3$ versus CP phase $\phi_a,\phi_b,\phi_c$ and Majorana neutrino mass $m_{N_1}$ where in (a) and (c) we set $m_{N_1}=300$ GeV. In (a),(b) we set $\sqrt{s}=1000$ GeV, $\left|R_{\mu_{N_i}}\right|^2=3\times10^{-4}$. In (c), (d) we set $\sqrt{s}=3000$ GeV, and we set $\left|R_{\mu_{N_i}}\right|^2=1\times10^{-5}$ in (c), $\left|R_{\mu_{N_i}}\right|^2=4\times10^{-6}$ in (d).}\label{fig6}
	\end{center}
\end{figure}

Finally, the results of the total CP violation with the case that there are three heavy Majorana neutrinos are shown in Fig.~\ref{fig6}.
Fig.~\ref{fig6a} and Fig.~\ref{fig6c} show that the influence from $\phi_a$, $\phi_b$ and $\phi_c$ are different, and the CP violation caused by $\phi_b$ are obviously smaller than those from $\phi_a$ and $\phi_c$ when there are three heavy neutrinos, and in this situation, the maximum value of total CP violation is nearly the same with that when there are two heavy neutrinos but a little bit higher.

The results show that the feature of CP violation caused by heavy Majorana neutrinos in LNV processes which we studied in previous works~\cite{Lu:2021vzj,Lu:2022pvw,Lu:2022wsm} are different from those in the LNC processes, CP violation in LNC processes are obviously smaller but nonzero. In LNV processes, the CP violation is obviously independent on mass of heavy Majorana neutrinos $m_{N_1}$, but in LNC process, when there are more than one generation of heavy Majorana neutrino, the CP violation will be influenced by $m_{N_1}$, it will decrease as the $m_{N_1}$ goes up. In LNV processes, the CP violation exists when there are at least two generations of heavy Majorana neutrinos, but in LNC process, only one generation of heavy Majorana neutrino will cause nozero CP violation though it is small. For the case there are three heavy Majorana neutrinos, we search for the maximum value of CP violation when we running all CP phase $\phi_a,\phi_b,\phi_c$ from $-\pi$ to $+\pi$ where we take the interval of each point as $\pi/10$, and the maximum value of total CP violation will reach $3.82\times 10^{-3}$ at $\phi_a=-\pi/2,\phi_b=\pi/2,\phi_c=3/5\pi$ in ILC case, and it will reach $7.43\times10^{-4}$ at $\phi_a=-3/5\pi,\phi_b=\pi/5,\phi_c=\pi/2$ in MuC case.

\section{SEARCH FOR HEAVY MAJORANA NEUTRINOS AT FUTURE LEPTON COLLIDERS}\label{sec4}
After analysing the CP violation, now we are interested in the prospect of searching for heavy Majorana neutrinos at future lepton colliders, so we analyse the signals and backgrounds at 500 GeV, 1000 GeV, 3000 GeV ILC and 3000 GeV, 10 TeV MuC. In order to identify the isolated lepton or jet, we indentify isolated jets and leptons by angular separation, which can be defined as 

\begin{equation}
	\label{14}
	\Delta R_{ij}=\sqrt{\Delta\phi_{ij}^2+\Delta\eta_{ij}^2},
\end{equation}
where $\Delta\phi_{ij} (\Delta\eta_{ij})$ is the azimuthal angle (rapidity) difference of the corresponding particles. We apply some basic acceptance cuts (referred as cut-I)
\begin{equation}
	\label{15}
p_T^\ell>10~\text{GeV},~\left|\eta^\ell\right|<2.5,~p_T^j>20~\text{GeV},~ \left|\eta^j\right|<3.0,~0.4<\Delta R_{ij}<3.0
\end{equation}
In order to purify the signal, the missing transverse energy is required to satisfy (referred as cut-II)
\begin{equation}
	\label{16}
E\slash_T < 60 \text{~GeV}
\end{equation}

There are too much diagrams of backgrounds in the SM for this process, so we simulate all of them with the SM model by \textsf{MadGraph5\_aMC@NLO}, we also simulate the signals via \textsf{MadGraph5\_aMC@NLO} with the New Physics model generated by $\text{F}_{\text{EYN}}\text{R}_{\text{ULES}}$. The parton shower and hadronization are performed with \textsf{Pythia-8.2}~\cite{Sjostrand:2006za}. 
We also give another cut for our signal process, a $W$ boson will decay hadronically, and we can reconstructed it from the two jets ($j_1,j_2$). Their invariant mass should be closest to $m_W$. This leads to a new cut (referred as cut-III):
\begin{equation}
	\label{17}
	\left|M_{j_{1}j_{2}}-m_W\right|<50~\text{GeV},
\end{equation}
where $j_1,j_2$ are the two jets decayed by $W$ boson. 

\begin{figure}[!htbp]
	\begin{center}
		\subfigure[]{\label{fig7a}
			\includegraphics[width=0.45\textwidth]{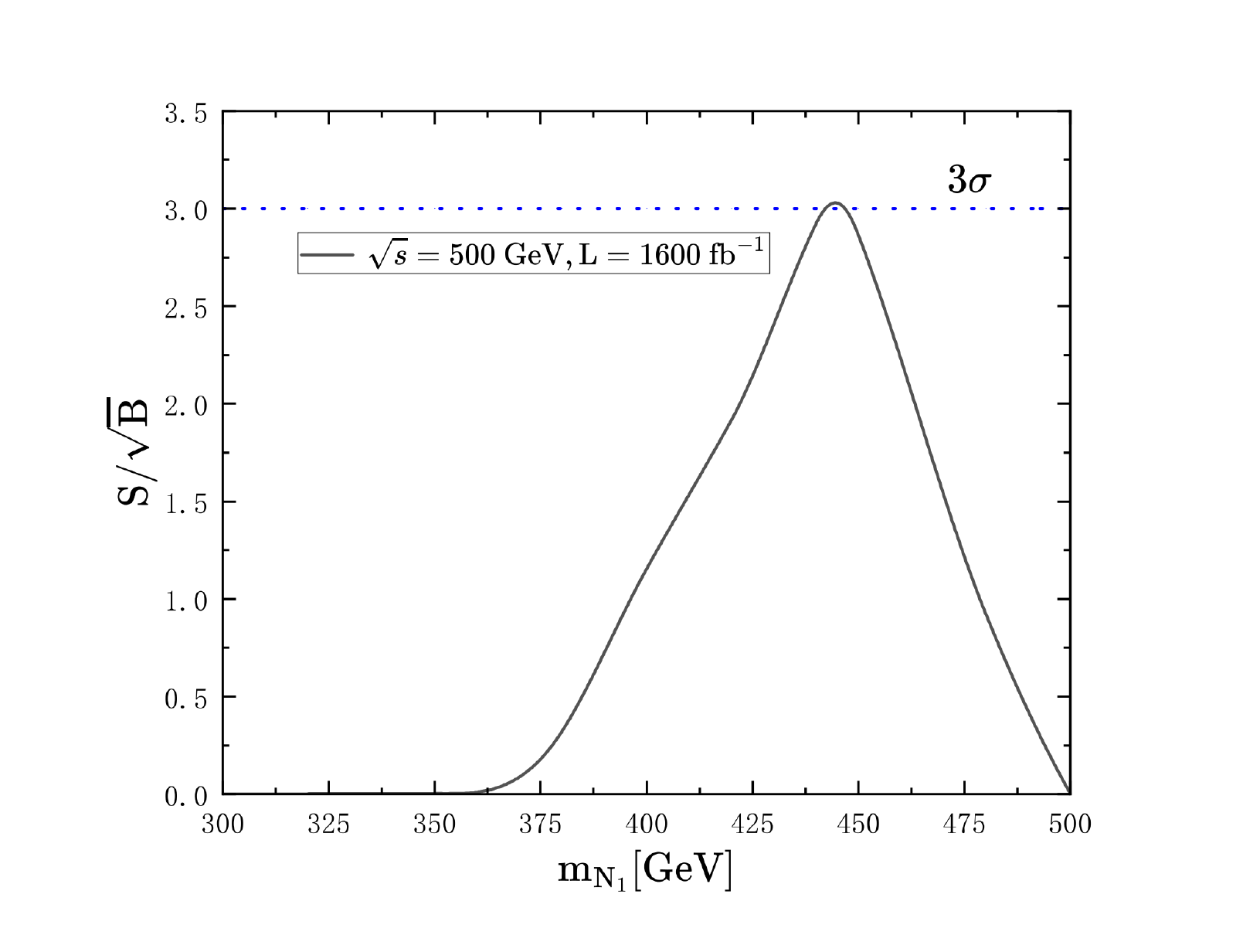} }
		\hspace{-0.5cm}~
		\subfigure[]{\label{fig7b}
			\includegraphics[width=0.45\textwidth]{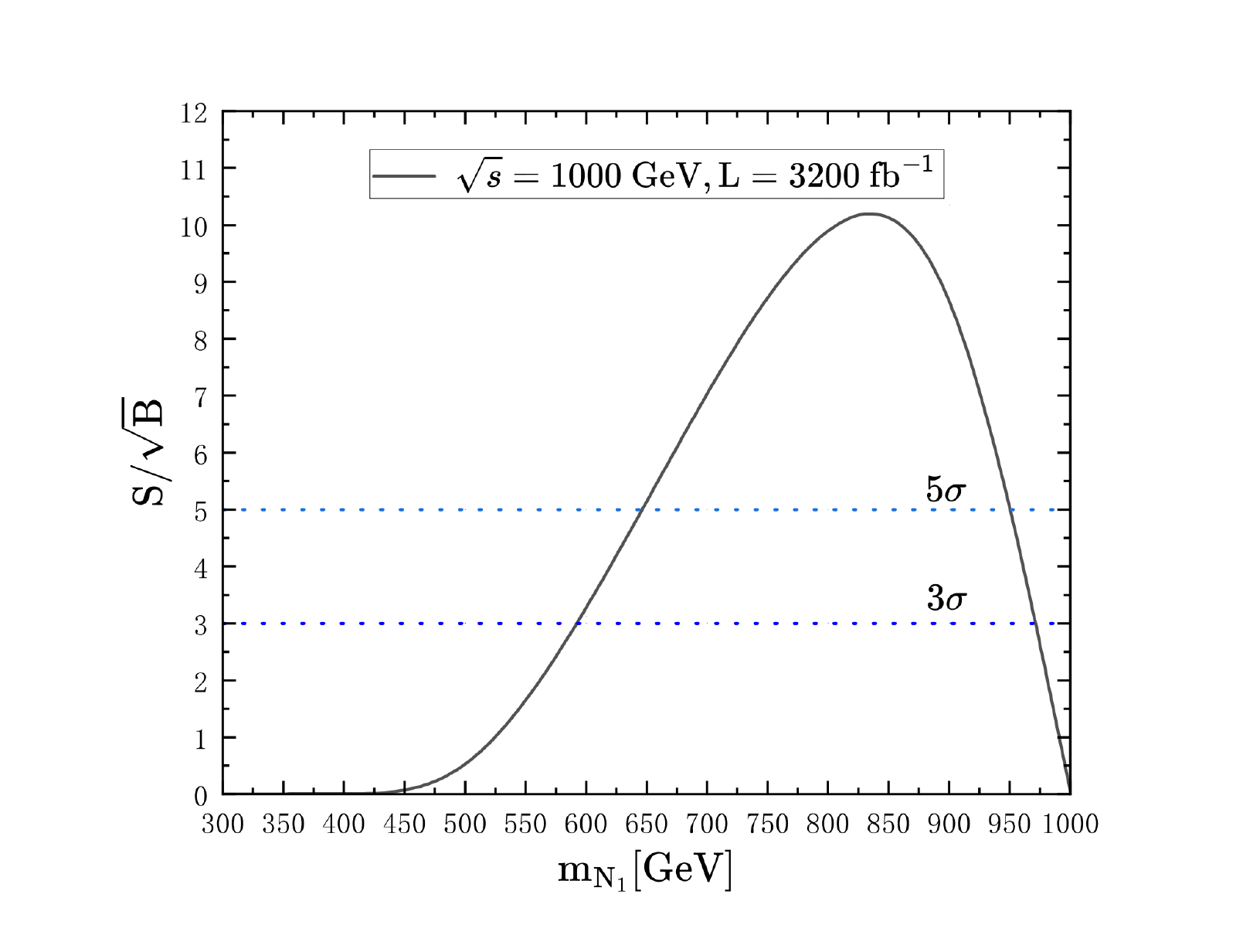} }
			\hspace{-0.5cm}~
		\subfigure[]{\label{fig7c}
	\includegraphics[width=0.45\textwidth]{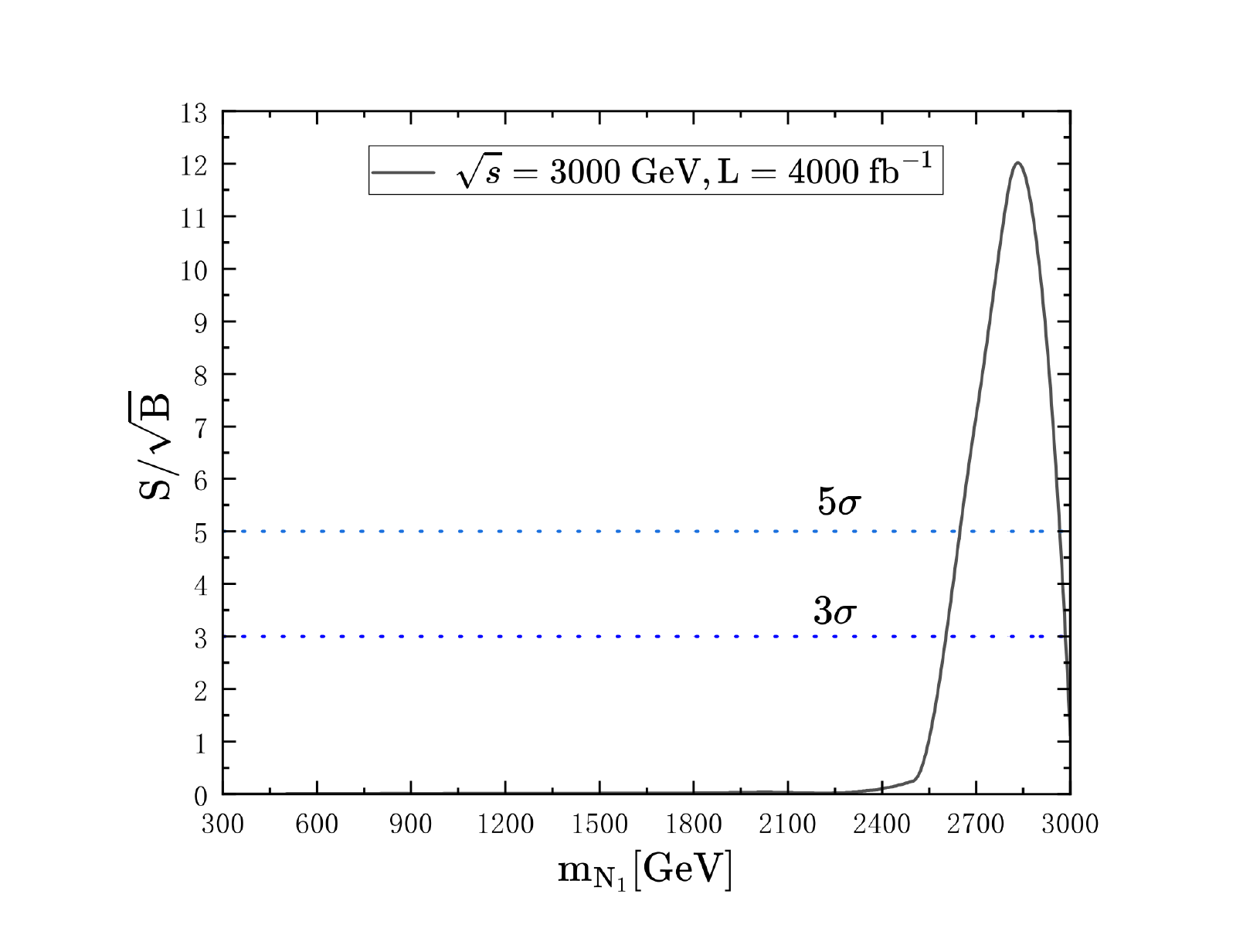} }
\hspace{-0.5cm}~			
		\caption{The statistical significance $S/\sqrt{B}$ as a function of $m_{N_1}$ with (a) the center-of-mass energy of $\sqrt{s}=500$ GeV, the integrated luminosity of ${\cal L}=1600$ fb$^{-1}$, (b) $\sqrt{s}=1000$ GeV, ${\cal L}=3200$ fb$^{-1}$, (c) $\sqrt{s}=3000$ GeV, ${\cal L}=4000$ fb$^{-1}$ at ILC, where all CP phases $\phi_a,\phi_b,\phi_c$ are set to zero.}\label{fig7}
	\end{center}
\end{figure}

In this process, for a better cut at background the transverse momentum of jets $\text{H}_{\text{T}}$ is required to satisfy (referred as cut-IV)
\begin{equation}
\label{18}
\text{H}_{\text{T}}>200~\text{GeV},
\end{equation}
After all the cuts on the backgrounds and signals, we get the results of statistical significance S/$\sqrt{B}$ as function of $m_{N_1}$ at 500 GeV, 1000 GeV in Fig.~\ref{fig7a} and Fig.~\ref{fig7b}. Here, we set all phases to zero. 

It is clear that at 500 GeV ILC, a $3\sigma$ discovery can be made for near 440 GeV $\leq m_{N_1}\leq$ 450 GeV with ${\cal L}=1600$ fb$^{-1}$, at 1000 GeV ILC, a $5\sigma$ discovery can be made for near 650 GeV $\leq m_{N_1}\leq$ 950 GeV with ${\cal L}=3200$ fb$^{-1}$, a $5\sigma$ discovery can be made for near 2650 $\leq m_{N_1}\leq$ 2950 GeV with ${\cal L}=4000$ fb$^{-1}$. It's hard to have a $3\sigma$ discovery at 3 TeV, 10 TeV MuC, the results are too small that we don't put the results in this paper. Finally, we give the total cross sections for signal processes in Table.~\ref{table1}.

\begin{table}[!htbp]
	\caption{Total cross sections for $\ell^+\ell^-\rightarrow \bar{\nu_\ell}\ell^-q\bar{q}^\prime$ at 500 GeV, 1000 GeV, 3000 GeV ILC with the integrated luminosity of $1.6$ab$^{-1}, 3.2$ab$^{-1}, 4$ab$^{-1}$ respectively and at 3000 GeV, 10 TeV MuC with the integrated luminosity of $1$ab$^{-1}, 10$ab$^{-1}$ respectively where all the CP phases $\phi_a,\phi_b,\phi_c$ are set to zero. }\label{table1}
	\begin{tabular}{|l|c|c|}
		\hline
		& \multicolumn{2}{c|}{$\sigma_{total}[\text{fb}]$ at ILC}\\
		\hline
		& ${\cal L}=1.6 \text{ab}^{-1}$,$\sqrt{s}=500 $ GeV & 3.48 \\
		\cline{2-3}
		$e^+e^-\rightarrow\bar{\nu_e}e^-q\bar{q}^\prime$& ${\cal L}=3.2 \text{ab}^{-1}$,$\sqrt{s}=1000 $ GeV & 5.47 \\
		\cline{2-3}	
		& ${\cal L}=4 \text{ab}^{-1}$,$\sqrt{s}=3000 $ GeV& 6.16\\
		\hline
		& ${\cal L}=1 \text{ab}^{-1}$,$\sqrt{s}=3000 $ GeV & $1.01\times10^{-2}$ \\
		\cline{2-3}
		\multirow{-2}{*}{$\mu^+\mu^-\rightarrow\bar{\nu_\mu}\mu^-q\bar{q}^\prime$}& ${\cal L}=10 \text{ab}^{-1}$,$\sqrt{s}=10 $ TeV & 0.241\\
		\hline
	\end{tabular}
\end{table}

	\section{SUMMARY}\label{sec5}
The small neutrino masses show that we need to expand the SM for explaining the tiny neutrino masses. An interesting model is the type-I seesaw mechanism which introduced heavy Majorana neutrinos that can lead to the CP violation in LNV process, the CP violation can give a new source to explain the baryon asymmetry in the Universe via leptogenesis. We have studied the heavy Majorana neutrinos in LNV process and there shold be at least two generations of heavy Majorana neutrinos to generate CP violation by the interference of contributions from different heavy Majorana neutrinos.

 In this work, we investigate the heavy Majorana neutrinos in an interesting LNC process, and find that only one generation of heavy Majorana neutrino can lead to CP violation, and several features of CP violation in the LNC process are different from that in LNV process. The CP violation in the LNC process is caused by the interference of contributions from different generations of heavy Majorana neutrinos and from the interference of contribution of the $s$-channel diagram and $t$-channel diagram. We consider three generations of heavy Majorana neutrinos $N_1,N_2,N_3$, and the possibility for searching it at future lepton colliders is studied at future lepton colliders 500,1000,3000 GeV ILC, and 3000 GeV,10 TeV MuC. The results show that there are great chances to explore the CP violation effects at future lepton colliders, and the exploring of CP violation can be a probe for studying the underlying new physics.

	\section*{ACKNOWLEDGEMENTS}
Z. Wang, X. H. Yang and X. Y. Zhang thank the members of the Institute of theoretical physics of Shandong University for their helpful discussions. This work is supported in part by National Natural Science Foundation of China (Grants No. 12235008, 12305106) and Natural Science Foundation of Shandong Province (Grant No. ZR2021QA040).

\end{document}